\documentclass[aps,prb,twocolumn,floatfix,showpacs,superscriptaddress]{revtex4}
\usepackage{amsmath}
\usepackage[dvips]{graphicx}
\usepackage{color}

\begin{document}
\newcommand{\gae}{$\Gamma_2(E)$}
\newcommand{\gaer}{$\Gamma_{2r}(E)$}
\newcommand{\gaz}{$\Gamma_2(0)$}
\newcommand{\h}{$h$}
\newcommand{\hc}{$h_c$}
\newcommand{\tf}{$t_h$}
\newcommand{\tb}{$t_\beta$}
\newcommand{\td}{$t_\Delta$}
\newcommand{\qea}{$q_{\textrm{EA}}$}
\def\sb{\mbox{\boldmath $\sigma$}}

\date{\today}

\title{Weak coupling study of decoherence of a qubit in disordered magnetic environments}
\author{E. A. Winograd}
\affiliation{Departamento de F\'{\i}sica, FCEN, Universidad de Buenos Aires,
Ciudad Universitaria Pab.I, (1428) Buenos Aires, Argentina.}
\author{M.J. Rozenberg}
\affiliation{Departamento de F\'{\i}sica, FCEN, Universidad de Buenos Aires,
Ciudad Universitaria Pab.I, (1428) Buenos Aires, Argentina.}
\affiliation{Laboratoire de Physique des Solides, CNRS-UMR8502, Universit\'e de Paris-Sud,
Orsay 91405, France.}
\author{R. Chitra}
\affiliation{Laboratoire de Physique Theorique de la Mati\`ere 
Condense\'e, UMR 7600, Universit\'e de Pierre et Marie Curie, Jussieu, Paris-75005, France.}

\date{}

\begin{abstract}
We study the decoherence of a qubit  weakly coupled to  frustrated spinbaths.  
We focus  on  spin-baths described by the classical Ising spin glass and
the quantum random transverse Ising model which are known to have complex  thermodynamic phase
diagrams as a function of   an external magnetic field and temperature.  Using a combination of
numerical and analytical methods, we show that  for baths initally in thermal equilibrium, 
the resulting  decoherence is  highly sensitive to the nature of the coupling to the environment and is qualitatively different in different parts of the phase diagram. We find  an unexpected strong non-Markovian decay of the coherence when the random transverse Ising model bath is prepared in an initial state characterized by  a finite temperature paramagnet.
 This is contrary to the usual case of exponential decay (Markovian) expected for spin baths in finite temperature paramagnetic phases,  thereby illustrating the importance of the underlying  non-trivial
 dynamics of  interacting quantum spinbaths.
\end{abstract} 

\maketitle

\section{Introduction}
The understanding and control of the decoherence of small quantum  systems is
central to a lot of recent developments in the fields of nanotechnology and 
quantum computers \cite{1-ver,2-ver,3-ver,4-ver,5-ver}.
For example, the efficacy of qubits, the basic building block of a quantum computer, which can
be spin qubits, Josephson junction qubits, charge or flux qubits depends largely on the environments, or
baths, to which they are often weakly  coupled. The effect of environments on the coherence of the qubit has been
studied in various contexts with particular emphasis on bosonic baths \cite{caldeira-leggett,weiss}. 
In the past few years, the realization of solid state qubits in semiconducting
heterojunctions has  also resulted in the study of spin baths  constituted of  spins.
 At very  low temperatures, the decoherence engendered by a spin environment 
is expected to  dominate  the loss of the coherence arising from a coupling to phononic degrees of freedom.  

In the limit of weak coupling between the qubit and the bath, the effect of intrabath interactions have been explored earlier in
other works \cite{italian-meanfield, lages, CC-prb, rossini, CC-prl,sib,ref1-2}.  
In some cases, interactions were found to decrease the rate of decoherence, though the exact opposite was seen in 
cases where the spin bath was on the verge of a  standard magnetic quantum phase transition \cite{CC-prl}. 
On the other hand, the  qualitative nature of the decoherence i.e., whether  it is Markovian or non-Markovian depends largely on the nature
of the initial state of the bath\cite{ ref1-1,ref1-3,ref1-2,deng}. 
For baths at zero temperature,  the decoherence is often non-Markovian.
 Examples include   the  well known spin-boson model \cite{caldeira-leggett}, 
  and models of interacting spin environments that we
already mentioned above \cite{CC-prb,CC-prl}. It has also been found in the case
of a  qubit interacting via the Fermi contact hyperfine coupling with a bath of polarized nuclear spins \cite{ref1-1}.  
However, for  baths in
thermal equilibrium  i.e., at {\em finite temperatures},  the decoherence  in the weak coupling limit is  expected to be  Markovian as
in the case of the spin-boson model.  This  aspect was also seen in the case of  interacting
spin baths at finite temperatures\cite{italian-meanfield, lages, CC-prb, rossini, CC-prl}.   The associated
Markovian decay rate  was also found to increase with temperature and, unlike the case of bosons, was found to saturate at high
enough temperatures. 

Since interactions between spins in condensed matter systems  are known to generate a whole
range of complex thermodynamic behaviours, it is interesting to ask whether interactions between the
bath spins generate novel behaviour for  the coherence of a qubit coupled to such a bath. In this general context,  various questions arise  naturally:  does the decoherence contain clear signals about the
underlying thermodynamic phase of the bath?
Is Markovian decay to be expected for all interacting spin environments at {\em finite temperatures}, where 
the bath is in a statistical mixture at temperature $T$, and not in a pure state \cite{CC-prb,pb,lpm}? And can decoherence or others physical quantities asociated with the qubit be used as a sensitive probe of the  dynamics of the underlying  spin environment?

Here, we shall explore some of these issues by considering the problem of the decoherence of a qubit 
induced by weak coupling to a model spin-bath with non-trivial dynamics arising from strong  frustration and prepared in an initial state which is in thermal equilibrium.  Highly frustrated  or disordered baths
were partially explored 
in [\onlinecite{CC-prb,lages}] where complete predictions for the decoherence could be made only for the
case of a one dimensional Ising bath or for spin shards in the infinite temperature limit.  Though these works illustrate the potential richness of interacting environments 
they also highlight the difficulty and limitations of using analytical  methods to study these problems.
 We adopt a disordered spin bath 
described by the meanfield random Ising model in a transverse magnetic field. This environment is characterized by
a rich thermodynamic  phase diagram which shows a spin-glass to paramagnetic transition as a function of temperature
and external magnetic field \cite{Hertz-Fisher}. 
The total Hamiltonian for the system is given by 
 \begin{eqnarray}\label{eq:ham}
H &= & H_B + H_{SB} \nonumber \\
H_B &=& \sum_{i < j}^NJ_{ij}\sigma_i^x\sigma_j^x+h\sum_{i=1}^N\sigma_i^z \nonumber \\
H_{SB}&=& \sigma_c^{a}\sum_i^N\lambda_i\sigma_i^{a}
\end{eqnarray}
where $H_B$ is the bath hamiltonian and $H_{SB}$ denotes the spin-bath interaction. The Pauli matrices  $\sb_c$ and
$\sb_i$ denote the spin of the qubit and  the  $i$-th spin of the bath 
respectively.
The  exchange energies  $J_{ij}$  are quenched random variables with a probability  distribution $P(J)=\frac{1}{\sqrt{2\pi}\Delta}e^{-J^2/2\Delta^2}$, where the standard deviation of the distribution $\Delta$
sets the unit of energy.   The qubit couples to the bath operator $V^a=\sum_i^N\lambda_i\sigma_i^{a}$, with $a$ representing either  the $z$ or the $x$ component of the spins.  
The coupling constants $\lambda$ are also 
taken to be quenched random variables. The physics of the bath is described 
by $H_B$, the random transverse Ising model (RTIM) which is known to have the following phase diagram:
at zero temperature,  for  transverse fields smaller than a critical field $h_c$, i.e., $h < h_c$, the system is in a 
spin glass phase \cite{marcelo-grempel}. For $h >h_c$ the system is magnetized with a gap in its
spectrum. At finite temperatures,  the spin glass order  survives
up to a critical temperature $T_{sg}(h)$\cite{marcelo-2,bdgz}. A physical system which is expected to be reasonably described by the above model is  $\textrm{LiHo}_{\textrm{x}}\textrm{Y}_{\textrm{1-x}}\textrm{F}_{\textrm{4}}$, where the Ho concentration, x, tunes
the system between different physical regimes \cite{rosenbaum}. It has been proposed as a minimal
toy model to investigate the effects of quantum entanglement of spins \cite{holmio}. 
The critical temperature $T_{sg}$, that sets the value of the coefficient $\Delta$, is of the order of $1\textrm{K}$ \cite{rosenbaum}, in the compounds.  Consequently, the  effective high temperature paramagnetic phases we study in this
paper correspond to rather low real temperatures, where  we expect the spin environment to 
dominate and  phononic contributions to the decoherence can be neglected.   There have also been propositions for engineering model magnetic  environments  with very small energy scales  using cold atoms \cite{rossini}.

We note that in the spin glass phase, only the
correlations involving the $x$-component of the spin  exhibit  spin glass
features  whereas the correlation of the $S_z$ components merely exhibit
a gap.
Clearly, it would be interesting to explore the effect of these phases  on the
decoherence of the qubit.  Moreover,  due to the inherent anisotropy of the bath, we expect the
decoherence to be  dependent on whether  the qubit couples to  the $x$ or $z$ components of the bath spins.  In this paper, we address these questions in the limit
of weak qubit-spinbath coupling.   Since this problem cannot be studied analytically, we use numerical exact diagonalization methods to calculate the bath
eigenstates and eigenvalues and consequently, the resulting decoherence. We obtain a rich spectrum of results for the coherence of the qubit and in particular, a non-Markovian  i.e., a non-exponential
decay of the coherence at finite temperatures.

The paper is organized as follows: in Sec.~\ref{sec:formalism}, we present  the weak coupling
formalism used to calculate the decoherence of the qubit.  We present our numerical method in  Sec.~\ref{sec:numerical}  and use it to study
the analytically tractable case of decoherence induced by an Ising chain spin-bath so as to benchmark our method. We compare our results for the Markovian decoherence rate  with known analytical results for an infinitely long chain, so to establish the importance of the finite size effects introduced by our numerical method. In Sec. \ref{sec:results}, we present
our results for the decoherence in the weak coupling regime for the long range Sherrington-Kirkpatrick model which has an Ising
spin glass phase at low enough temperatures and present a comparison of our
results with some analytical results which are known in the high temperature paramagnetic phase.
In Sec.~\ref{sec:TIM},  we present in detail our main results for the decoherence induced
by the RTIM spin-bath
for two different couplings of the qubit to the bath.
We conclude with a discussion of our results in Sec. \ref{sec:conclusions}.

\section{Weak coupling formalism} \label{sec:formalism}
In this section, we   summarize the weak coupling formalism  \cite{CC-prb}   used to calculate the decoherence. This approach
is valid provided the energy associated with the qubit-bath coupling is smaller than all the scales of the
bath.  
We  first assume that at time $t=0$, the combined system of the  central spin and the bath have  a factorizable   initial density matrix: $\Omega=\rho(0)  \otimes \rho_B$. 
Depending on the coupling of the  central spin to the bath operator $V^a$, the qubit 
  is in a pure state whose basis vectors are defined by 
$ |\psi^a \rangle = \alpha|\! \gets^a \rangle + \beta|\! \to^a \rangle$ ($\rho(0) = |\psi^a \rangle \langle \psi^a |$)  where
$|\!\gets^x\rangle =|\!\gets\rangle $, $|\!\to^x\rangle= |\!\to \rangle$
and $|\!\gets^z\rangle=|\!\uparrow \rangle$, $|\!\to^z\rangle=|\!\downarrow \rangle$.
The   bath is chosen to be at thermal  equilibrium with temperature $T\equiv 1/\beta$\, leading to a density
matrix
\begin{equation}
\rho_B = \frac{e^{-\beta H_B}}{Z} \label{eq:rhoB}
\end{equation} 
where $Z=\mathrm{Tr} \, \exp(-\beta H_B)$ is the bath partition function. 
The time evolved reduced density matrix  is given by
\begin{widetext}
\begin{equation}
\rho(t)= |\alpha|^2 \vert \!\gets^a \rangle \langle \gets^a \! \vert + 
|\beta|^2 \vert \! \to^a \rangle \langle \to^a \! \vert + 
M(t) \alpha^* \beta \vert \! \to^a \rangle \langle \gets^a \! \vert 
+ M(t)^* \alpha \beta^* \vert \! \gets^a \rangle \langle \to^a \! \vert
\end{equation}
\end{widetext}
where the factor
\begin{equation}
M(t)= \mathrm{Tr} \left( e^{-i(H_B+V^a)t} \;\rho_B\; e^{i(H_B-V^a)t} \right) \label{eq:Mt}
\end{equation}
is a measure of the decoherence induced by the bath at time $t$\cite{weiss,CC-prb}. 
Note that $\mathrm{Tr}$ denotes the usual trace as $H_B$ and $V^a$ are operators in the bath Hilbert space. 
For weak coupling to the environment i.e., for small $\lambda$, we can use the super-operator formalism\cite{weiss} to obtain the following form for the decoherence\cite{CC-prb}, being the Laplace transform of $M(t)$ 
\begin{equation}
{\tilde M}(z)= -i \int_0^\infty dt \, e^{izt} M(t)  \label{eq:LTdef} 
\end{equation}  
where  $z$ is a complex variable with $\mathrm{Im} z >0$. 
As shown in [\onlinecite{CC-prb}], this Laplace transform can be written as 
\begin{equation}
{\tilde M}(z)= \left[ z-\Sigma(z) \right]^{-1} \label{eq:LT} 
\end{equation}  
\noindent
where the self-energy $\Sigma$  up to second order is given by 
\begin{eqnarray} 
\Sigma_2(z) & = & 2\mathrm{Tr} (V^a \rho_B ) - \nonumber \\
& & 2i \int_0^\infty dt\, e^{izt} \left[ \langle V^a(t)V^a\rangle_c + \langle V^aV^a(t)\rangle_c \right] \label{eq:soSigma}
\end{eqnarray}       
where the connected correlation functions are defined as
\begin{equation} 
 \langle V^a(t)V^a\rangle_c = \langle V^aV^a(t)\rangle -\langle V^a(t) \rangle  \langle V^a \rangle \label{eq:conncorr}
\end{equation}   
The coherence $M$ can thus be written 
in terms of the real functions $\Lambda_2$ and $\Gamma_2$ defined by
\begin{equation}
\Lambda_2 (E)-i\Gamma_2(E)= \lim_{\eta {\to} 0^+} \Sigma_2(E+i\eta) \label{DeltaGamma}
\end{equation} 
where $E$ is real.
This then leads to the weak coupling result 
 \begin{equation}
\Theta(t) M(t) = \frac{i}{2\pi} \int dE \frac{e^{-itE}}{E-\Lambda_2(E)+i\Gamma_2(E)} \label{eq:ThetaM} 
\end{equation}     
where $\Theta(t)$ is the Heaviside step function and $\Lambda_2$ and $\Gamma_2$ 
satisfy standard Kramers-Kronig relations.
In general, $\Gamma_2(E)$ might have analytic as well as non-analytic parts.  It is interesting to note that in the weak coupling limit,  the decoherence is essentially
dictated by  \gae\  which  as can be seen from (\ref{eq:soSigma}) is proportional to the symmetrized dynamic structure factor of
the bath.

Typically for spin baths, we encounter situations where, one can have a net magnetization along a certain spin direction  $\langle V^a \rangle = \kappa  \neq 0$, and/or  magnetic ordering
where 
   $ \langle V^a(t)V^a\rangle_c = \mu + f(t)$  with $\kappa$ and $\mu$ being constants independent of time, and $f(t)$ a time-varying function.
The resulting self energy in the Laplacian variables then takes the form
\begin{equation} 
\Sigma_2(z)=2\kappa + 4 \frac{\mu}{z} + g(z) 
\end{equation}
where $g(z)$ is some analytic  function of $z$.
For the asymptotic decoherence,   the first two terms  generate oscillations resulting in
a decoherence of the form
\begin{equation}\label{eq:mt-osc}
M(t)= \exp{(2i\kappa t)} \cos{(2\sqrt{\mu}t)}  {\tilde M}(t)
\end{equation}
where \cite{CC-prb}
\begin{equation}\label{eq:mtilde}
\ln {\tilde M}(t) \simeq - \frac{2}{\pi} \int dE\,  \frac{\sin (tE/2)^2}{E^2} \Gamma_2(E) \quad .
\end{equation} 
and
\begin{equation}
\Gamma_2(E)= [{\tilde f}(E) + {\tilde f}(-E)] 
\end{equation}
being $\tilde f$ the Fourier transform of the function $f(t)$. (\ref{eq:mtilde}) is also
known as the time-convolutionless projection operator (TCL) approximation \cite{royer}.  The true intermediate time decoherence is
expected to be lightly modified with respect to the result predicted by \eqref{eq:mtilde}, but it is expected
to be qualitatively similar to that predicted by \eqref{eq:ThetaM}, which is valid for all $t$. 
In the absence of any intrinsic dynamics of the central spin, the second order aproximation used here, leads to an equation for the decoherence which resembles that obtained for the spin-boson model \cite{arXiv}. Though  \eqref{eq:ThetaM} is exact for the spin-boson case,  
here it is valid only in the limit of weak qubit-spinbath coupling.  Morevoer,  not all spectral densities \gae\ obtained from interacting	systems can be simulated by the usual non-interacting boson baths.

For the disordered systems studied here, the correlation functions (\ref{eq:conncorr}) and (\ref{eq:soSigma}) 
need to be averaged over the probability distributions of both
the exchange interactions $J_{ij}$ and the coupling constants $\lambda_i$. 
 In the rest of the paper, the coupling constants $\lambda_i$  in \eqref{eq:ham} are chosen  to have the following disorder averages:  $\overline{\lambda_i }= 0$ and 
$\overline { \lambda_i \lambda_j} =(\lambda^2/N)\delta_{ij}$ where $N$ is the total number of spins in the bath. This choice leads to the vanishing of the first order correction to the
self energy  since the disorder average of $\kappa$ is zero. The disorder averaged $\Gamma_2$, is then directly related to the connected local spin correlation functions $\overline{\langle V^a V^a\rangle_c}$, and it is evaluated using the spectral representation
for the dynamical structure  factor
\begin{widetext}
\begin{eqnarray}\label{eq:spectral}
S^{aa}(E) & = & \int_{-\infty}^{\infty} dt  \exp{(iEt)} \overline{\langle V^a(t) V^a\rangle_c} \nonumber \\
&=&  \frac{2\pi \lambda^2}{NM} \sum_{m=1}^M {\frac{1}{Z^{(m)}}}\sum_{i=1}^N \sum_{j,k=1}^{2^N} \exp{(-\beta E_j^{(m)})}
|\langle j^{(m)}| S^a_i|k^{(m)}\rangle |^2 \times \delta(\omega - E_j^{(m)} + E_k^{(m)}) -2\pi  \lambda^2 m_a^2 \delta(E)  \nonumber \\
&=&  2\pi \mu \delta(E) + {\tilde f}(E)
\end{eqnarray}
\end{widetext}
where  $M$ is the number of realizations of disorder for the exchange interactions $J_{ij}$ and $ m_a^2$ is the disordered average of the square of the local magnetic moment.
Our method for evaluating  (\ref{eq:spectral})  is presented below.

\section{Numerical method}\label{sec:numerical}
To compute ${\Gamma}_2$ for the RTIM spin-bath
we use exact  diagonalization methods. The advantage of the full diagonalization is that it permits us to calculate finite temperature quantities as well. 
We  first study the coherence of the qubit coupled to a random Ising chain bath described $H_B$ with the external 
field set to $h=0$ , where known analytical results
can be used to benchmark the
implemented numerical method. The interactions
in the bath are confined to nearest neighbors on the chain and the spin-bath coupling is taken to be in the $a=z$ direction.

\begin{figure}[bpt]
\centering
\includegraphics[scale=0.3,angle=0]{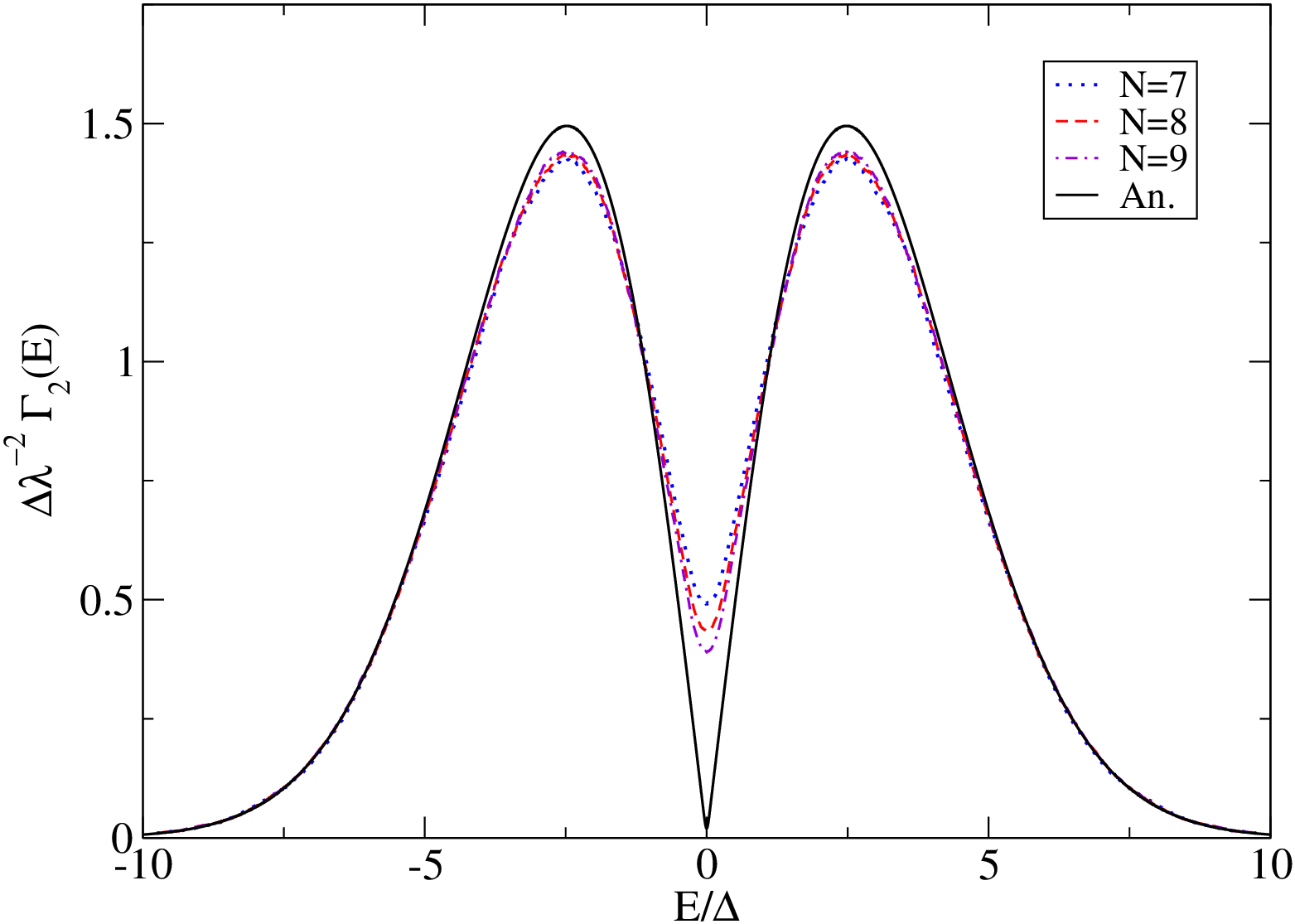}
\caption{\label{fig:sizeeffects}(Color online) $\Delta\lambda^{-2}\Gamma_2(E)$ as a function of $E$ for $T=0.01\Delta$ and 
different $N$. Note that as $N$ increases, the numerical results approach the analytical
result (An.). A
similar result is found for the SK long range system.}
\end{figure}

 We consider baths with the number of spins $N$ varying from  3 to 10 spins.  For a given set of exchange interactions,  the resulting
hamiltonian which is a $2^N\times 2^N$ matrix   is exactly diagonalized and the eigenenergies and functions are obtained.  These are then used to compute the dynamical structure factor  and
$\Gamma_2(E)$. Since the interactions are quenched random variables, we repeat the above
procedure for 
 several thousands of 
realization of disorder  and we obtain the disorder averaged $\Gamma_2$.
In Fig.~\ref{fig:sizeeffects}, we  plot $\Gamma_2(E)$ for  the case of the Ising bath 
for different sizes of the bath, and fixed temperature $T=0.01\Delta$.  The analytical results
for the thermodynamic case are also shown in the graph.

\begin{figure}[bt]
\centering
\includegraphics[scale=0.3,angle=0]{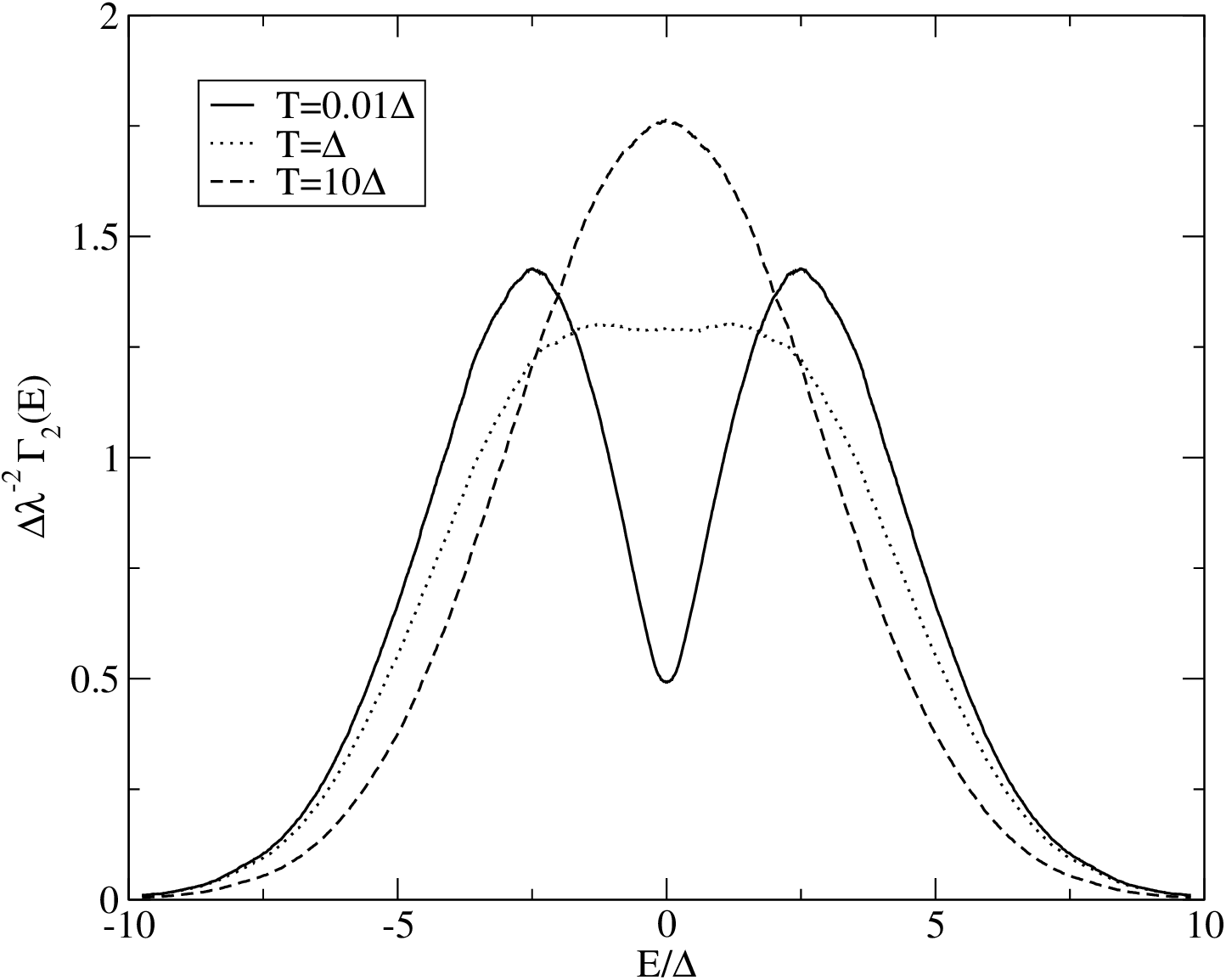}
\caption{\label{fig:tempeffects}$\Delta\lambda^{-2}\Gamma_2(E)$ as a function of $E$ for different $T$
for $N=7$. }
\end{figure}

Note that as the number of spins $N$ in the chain increases, the numerical results  converge to the analytical
 curve  obtained for a thermodynamic bath. At high $E$, $\Gamma_2(E)$ is  almost independent of 
the size of the bath. This is due to the fact that the typical energy 
bonds are of order $J_{ij}$ ($\mathcal{O}(\Delta)$) (independent of the
 size of the bath), which are sufficiently small to produce any modification in the high energy spectrum.
At low energies,  size effects are most significant, since at finite sizes there are no exact cancellations that lead to the vanishing of \gae\ at $E=0$. Nevertheless, the spectra shows good uniform convergence as $N$ increases.
At higher $T$, when the thermal energy is higher than the typical energy of the bonds, the convergence is even better and 
\gae\ is  roughly independent of the size of the bath.
In Fig. \ref{fig:tempeffects}, we plot the variation of \gae\ for different temperatures.
As shown in Ref. \onlinecite{CC-prb}, it is the low frequency part that is mainly affected by the temperature effects.

At high $T$, $\Gamma_2(E)$ is basically a broad peak of width $\Delta$, centered at 
$E=0$. This is due to the thermal fluctuations which result in uncorrelated spins,  and 
$\Gamma_2(E)$ merely  reflects the distribution of the bonds $J_{ij}$ and the structure of the energy levels.
Both temperature and size effects are seen to  have an impact on the low frequency part of \gae\ with the consequence that
the asymptotic decoherence is more sensitive to thermal fluctuations and finite size effects.

We use our numerical results for \gae\ to calculate the decoherence  $M(t)$ given by
\eqref{eq:mtilde}.  A comparison of our results with the analytical results obtained for the
infinite chain \cite{CC-prb} permits us  to gauge the importance of  finite size effects.  As for the infinite chain case of Ref. \onlinecite{CC-prb},  the decoherence  for the finite chain can also be analyzed in terms of three regimes described by the two characteristic times:
 $t_{\Delta}=\Delta^{-1}$ and  $t_{\beta}=\beta=T^{-1}$. At short times, the coherence is given by the universal gaussian, 
$\ln{M(t)}=-t^2(2\pi)^{-1}\int{dE}\Gamma_2(E)$, given by the sum rule of \gae
\begin{equation}\label{eq:sumrule}
\int_{-\infty}^\infty dE\phantom{i} \Gamma_2(E)=4\pi\lambda^2/\Delta
\end{equation}

\begin{figure}[bt]
\centering
\includegraphics[scale=0.32,angle=0]{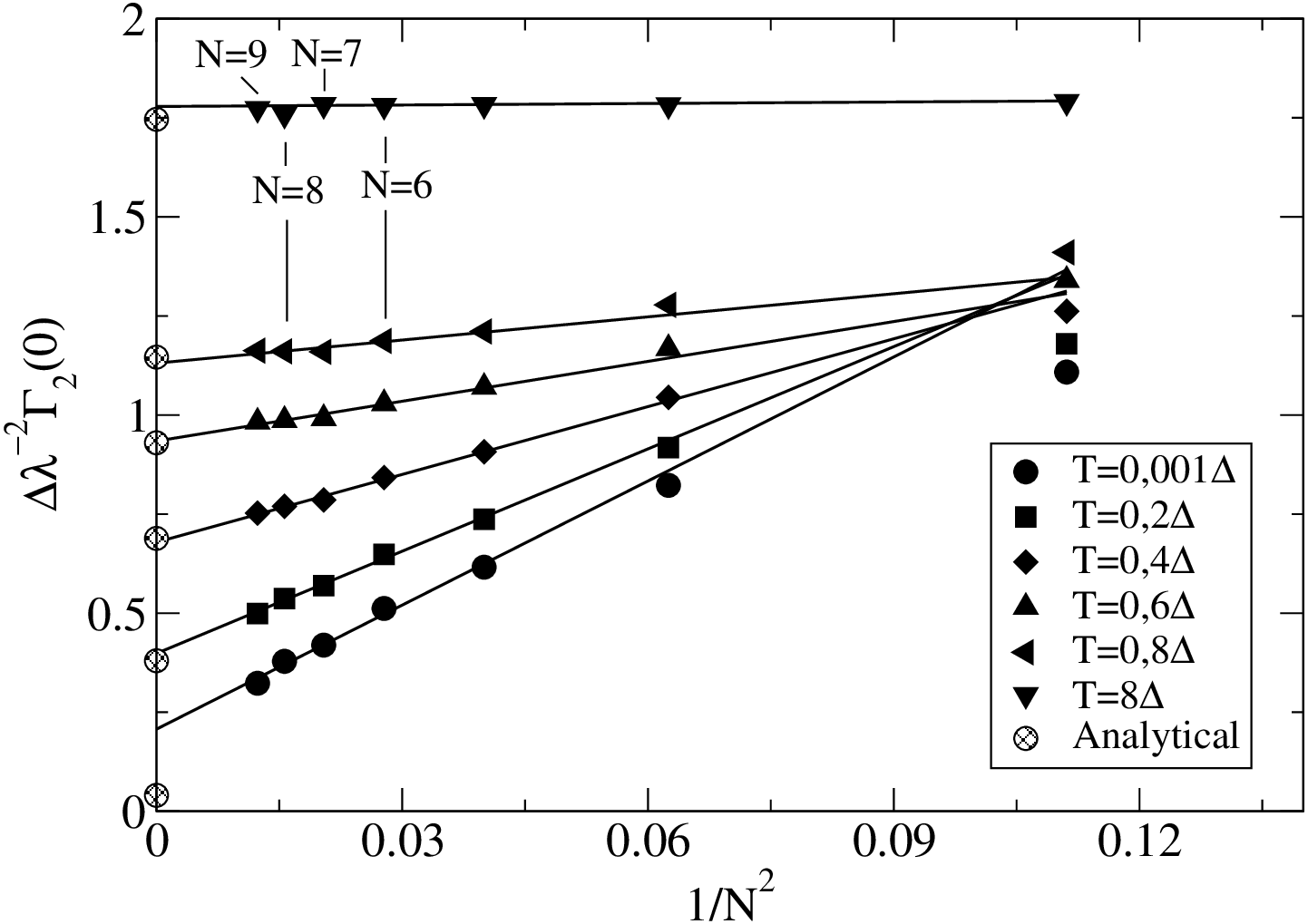}
\caption{\label{fig:cadgamma0}$\Delta\lambda^{-2}\Gamma_2(0)$ 
as a function of $1/N^2$. Note the excellent agreement of the extrapolated  value of \gaz and the analytical one, except at low $T$ ($T\lesssim0.2\Delta$).}
\end{figure}

\begin{figure}[bt]
\centering
\includegraphics[scale=0.3,angle=0]{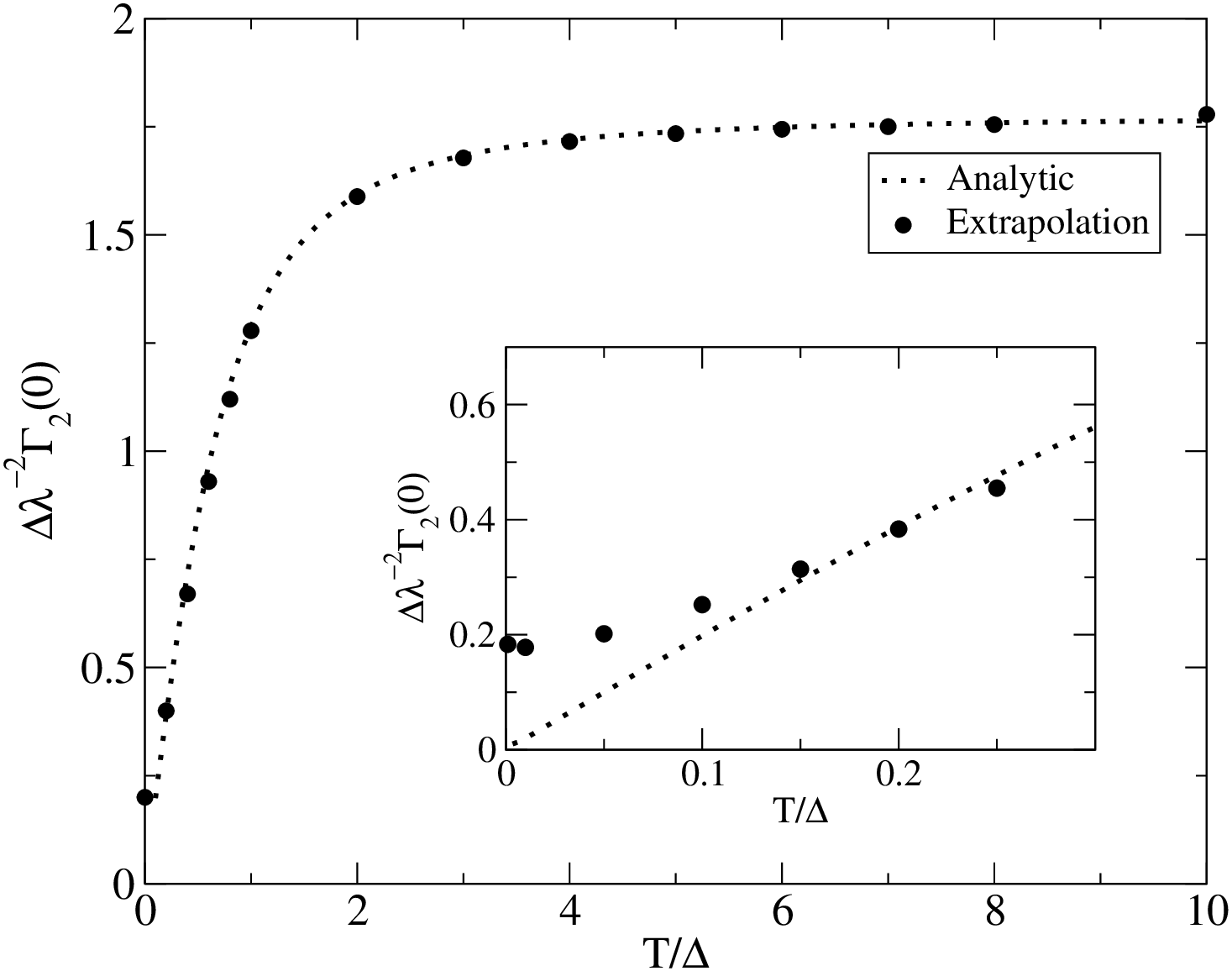}
\caption{\label{fig:cadgamma0T} $\Delta\lambda^{-2}\Gamma_2(0)$ (extrapolated to the
thermodynamic limit)
as a function of $T$. Good agreement with analytical results (dotted curve) except at $T\lesssim0.2\Delta$ (inset).}
\end{figure}

The asymptotic Markovian regime, where the coherence decays exponentially is characterized 
by the values of \gae\ at $E=0$ (for $t\rightarrow\infty$, $\ln{M(t)}\propto-\Gamma_2(0)t$).
At high $T$, $M(t)$ remains practically constant up to $t\simeq$\td, where a final Markovian regime arises. When $T\lesssim\Delta$, the 
gaussian regime is followed by an intermediate time regime where the coherence decay as a power law, before reaching the Markovian regime.
A careful inspection  of our numerical data shows that, \gaz\ scales as $1/N^2$ (Fig. \ref{fig:cadgamma0}).
The  extrapolated values of \gaz\ for $N\rightarrow\infty$ as a function of $T$ are plotted in Fig.~\ref{fig:cadgamma0T}, where it can be seen 
that there is an excellent agreement between the analytical values given by\cite{CC-prb} $\Gamma_2(0)=2\pi\lambda^2\int{dJ P(J)^2(1-\tanh(\beta{J})^2)}$, and the extrapolated values except for $T\lesssim0.2\Delta$ (inset of Fig. \ref{fig:cadgamma0T}). At very small $T$, since both size effects and 
temperature effects are non negligible at low energies, we surmise that the accessible system sizes are not sufficient to reach the scaling regime.

To summarize, we see that our numerical method does reproduce the 
expected analytical behavior for the spin chain bath except at low temperatures
where, the Markovian rate is overestimated in our approach. Nonetheless, our results
reproduce the existence of the different regimes. We would like to draw attention to the
fact that the benchmark case of the chain is in a sense a worst case scenario. In fact, for a 1-d spin-bath one expects to have stronger finite size effects
compared to the models to be studied in the rest of the paper: i.e., infinite-range models
which are known to have much reduced finite size effects \cite{marcelo-1,marcelo-2}.


Before leaving this numerical method section we would like to note that our method differs substantially from others
that are based on the solution of finite size clusters and cluster expansions. Those methods usually deal with a non-disordered
model hamiltonian, which is solved for a small system size. The finite size effects are dealt with by implementing various
cluster expansion schemes. In those cases, the finite nature of the cluster provides a discrete pole structure with a finite low
energy cut-off and hence a finite recursion time that provides a long time cut-off for the study of the decoherence effects.
In our case the situation is different. The effect of the bath enters through the spin susceptibility, and this quantity in
a disordered model is computed through the disorder average. Since we have a continuous (Gaussian) distribution of couplings,
the disorder-averaged susceptibility, though computed on finite size clusters, does not have a finite low-energy cut-off.
Therefore, the decoherence does not have a finite recursion time. All the systematic errors in our approach are due to the
second order approximation, which is safe so long the coupling between the qubit and the bath is small, and to the finite
size of the clusters that are diagonalized to compute the susceptibility of the bath's Hamiltonian. This latter quantity
can be, nevertheless, extrapolated to large system sizes, as we shall see latter. This is validated by the benchmark of 
the method against an analytically solvable case that we describe in the beginning of next section.

\section{Models of spins baths}
In this section, we apply our numerical method to study the decoherence induced by
interacting quantum spin-baths which do not have conventional magnetic order like
ferromagnetism or antiferromagnetism. The models studied all have infinite range interactions:
i.e., all spins interact with  each other  which leads to the presence of a significant geometric
frustration. Similar models have been studied in Ref. \onlinecite{lages}, where the focus was
on the nature of the distribution of the energy level spacings and their impact on the decoherence.
A standard lore is that thermal fluctuations  in the weak coupling regime always leads to 
a Markovian i.e., exponential decoherence. Here, we will discuss cases, where the decoherence
remains highly non-Markovian in certain finite temperature paramagnets.

\subsection{Infinite-ranged Ising bath with $h=0$}\label{sec:results}

\begin{figure}[bt]
\centering
\includegraphics[scale=0.3,angle=0]{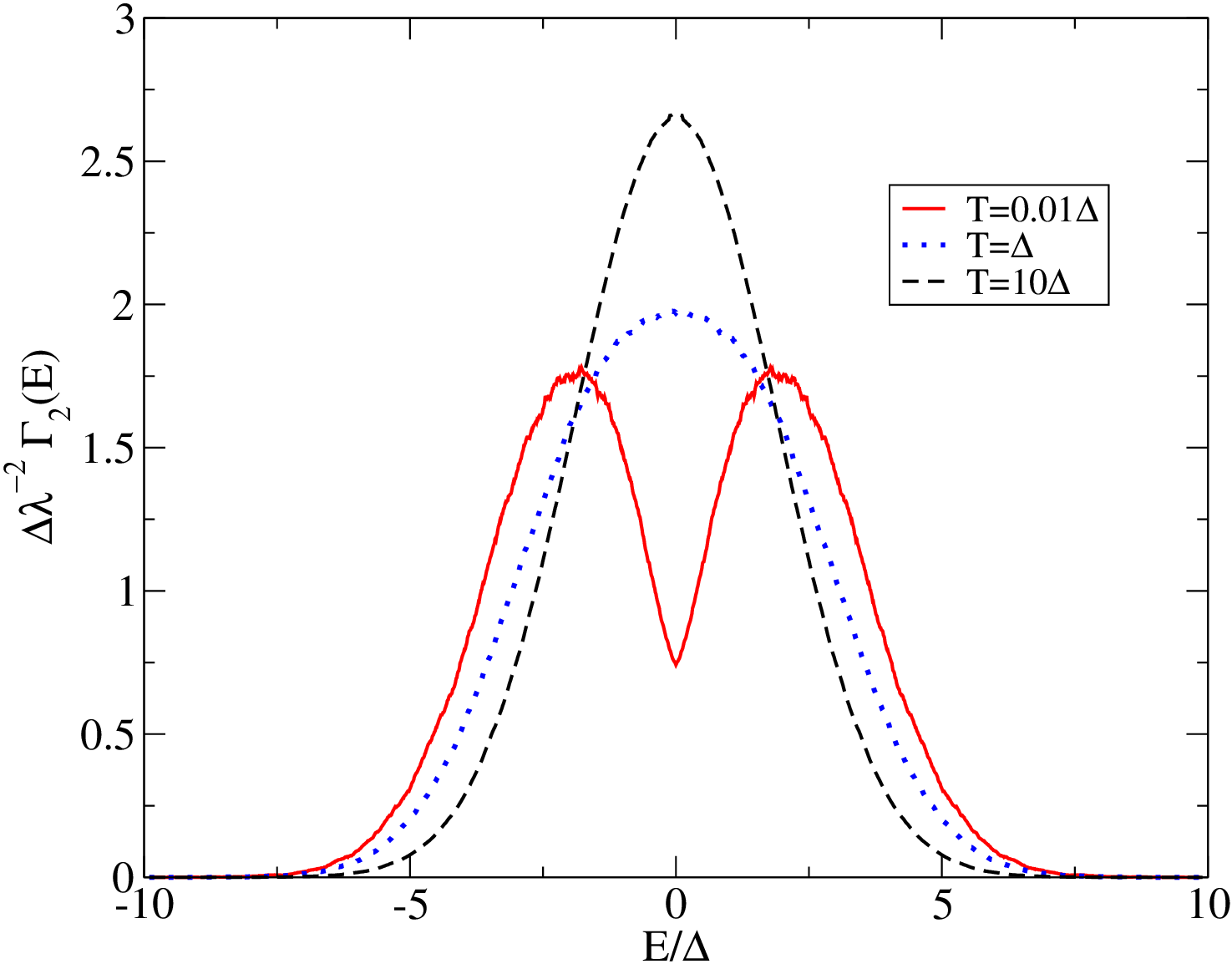}
\caption{\label{fig:csg97N9}(Color online) $\Delta\lambda^{-2}\Gamma_2(E)$ for different values of $T$ and $N=9$, for the
SK bath. The results are qualitatively similar   to that of the chain. }
\end{figure}

We study the decoherence induced by the infinite-range random Ising bath also known as the
Sherrington-Kirkpatrick (SK) model \cite{sk}.  For a gaussian distribution with zero mean of the
exchange interactions between the spins, the system is paramagnetic except below  the spin-glass transition temperature
$T\le T_{sg}=\Delta$  where the system develops spin glass order.  Note that for the infinite range model, extensiveness of the free energy requires as to scale the interactions $J_{ij} \to J_{ij}/N$.  In this case, there is no first order contribution to the self energy and the second order contribution is
given directly by the full correlation function as opposed to the connected correlation function of (\ref{eq:conncorr}) i.e., both $\kappa=\mu=0$. Moreover, $\Gamma_2$ is  proportional to the probability distribution of the local magnetic fields as discussed in Ref. \onlinecite{CC-prb}.  This distribution is
straightforward to evaluate analytically in the paramagnetic phase, but numerical methods are
required to obtain the same in the spin glass phase.   Earlier work only discussed the evolution of the Markovian rate close to the spin glass transition temperature \cite{CC-prb}. Here, we compute the  coherence  for all times and find an asymptotic Markovian regime at all
finite temperatures.
 Our results  are shown in Fig.~\ref{fig:csg97N9}
and we see that they qualitatively resemble the results obtained for the Ising chain. Moreover, our method reproduces known
analytical results in the paramagnetic phase and also approaches the expected linear behavior of $\Gamma$  at low frequencies as
$T \to 0$ \cite{sk}. 

\begin{figure}[bt]
\centering
\includegraphics[scale=0.32,angle=0]{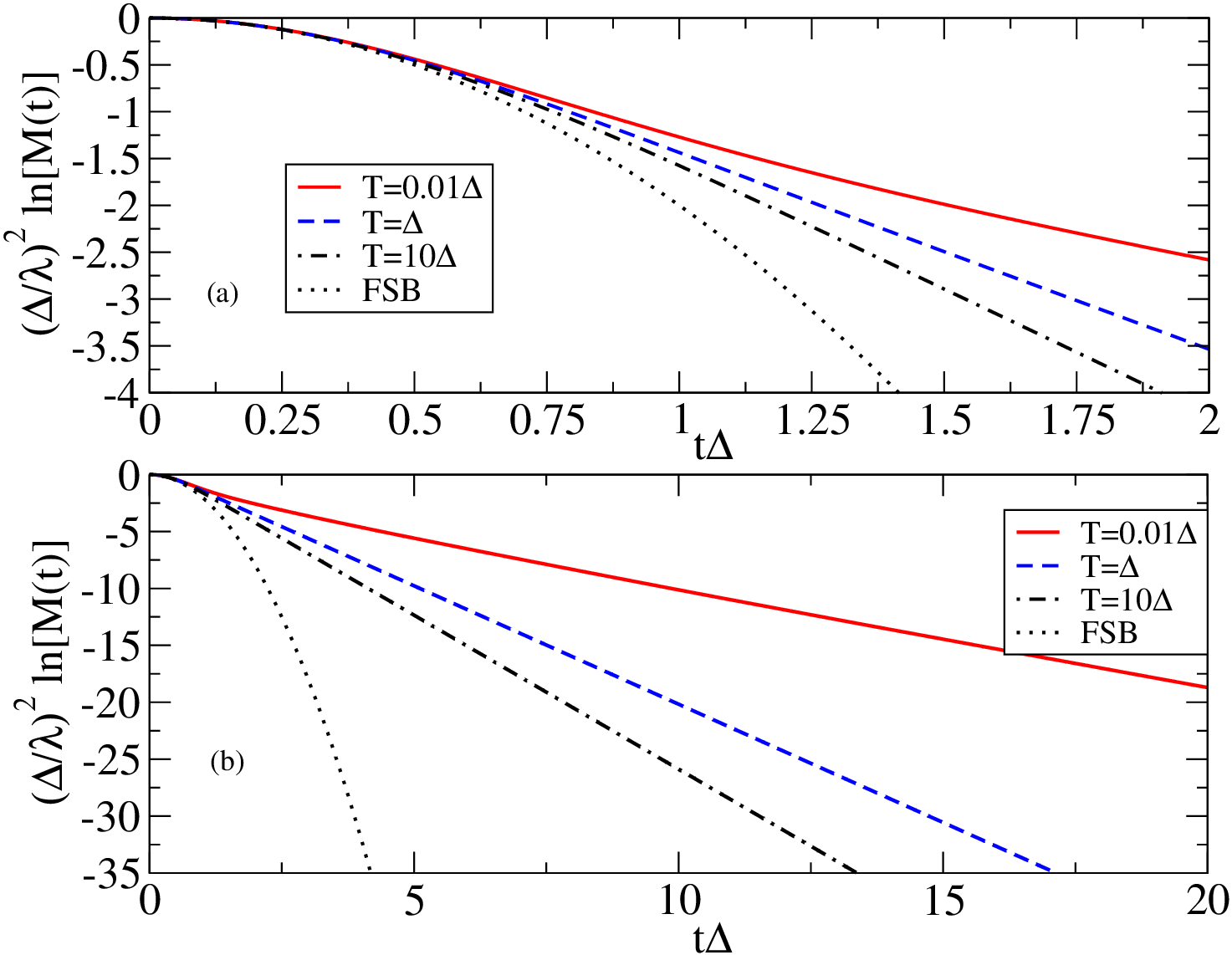}
\caption{\label{fig:csgMt}(Color online) $(\Delta/\lambda)^{2}\ln{M(t)}$ for the SK bath and the comparison with the free spin-bath (FSB); (a) for 
short times, where it can be seen that at $T=\Delta$, there's a change of the curvature in
$\ln{M(t)}$; (b) for long times.}
\end{figure}

\begin{figure}[bt]
\centering
\includegraphics[scale=0.3,angle=0]{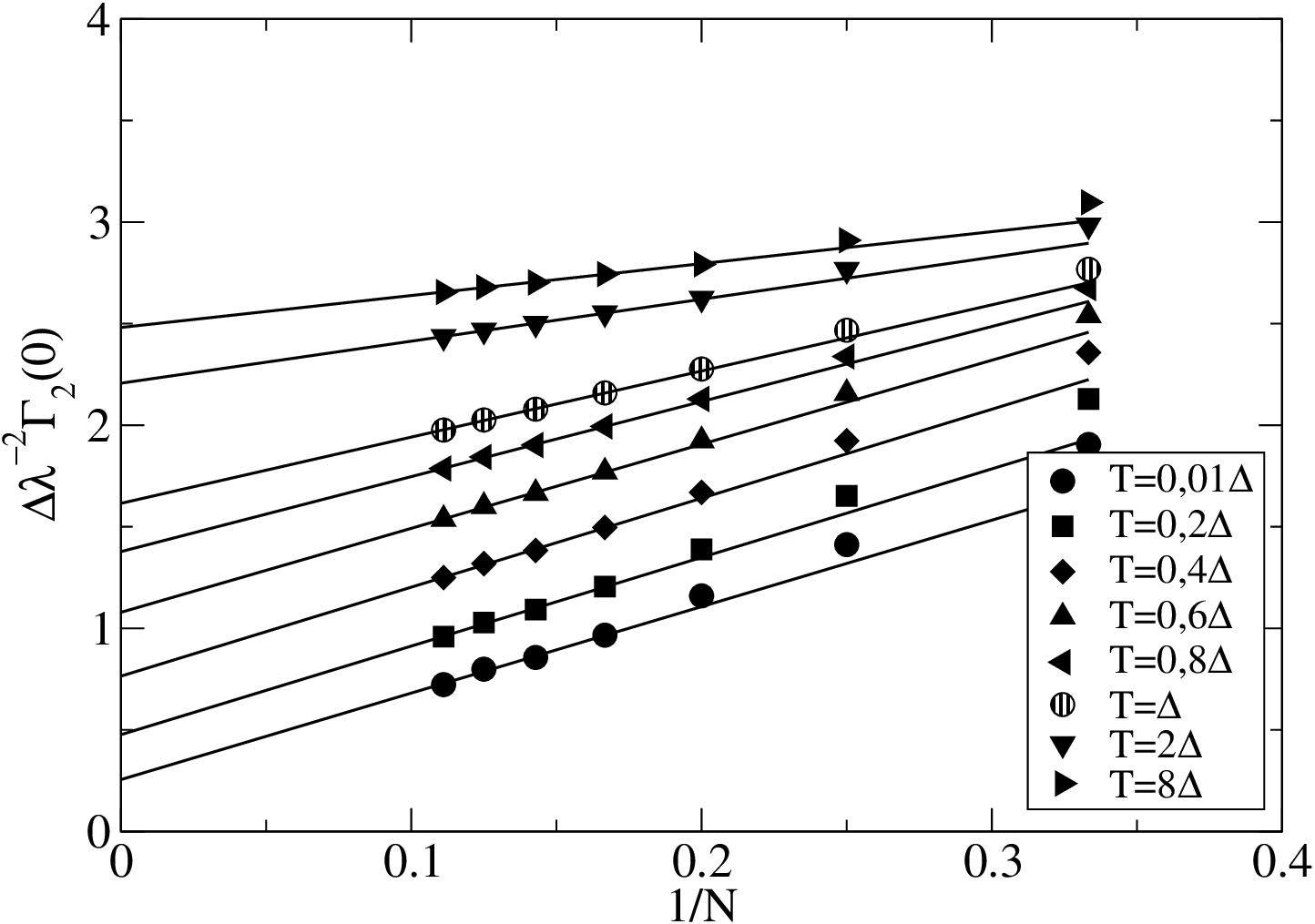}
\caption{\label{fig:csggamma0}Scaling plot for $\Delta\lambda^{-2}\Gamma_2(0)$ 
as a function of $1/N$ for the SK bath.}
\end{figure}

\begin{figure}[bt]
\centering
\includegraphics[scale=0.28,angle=0]{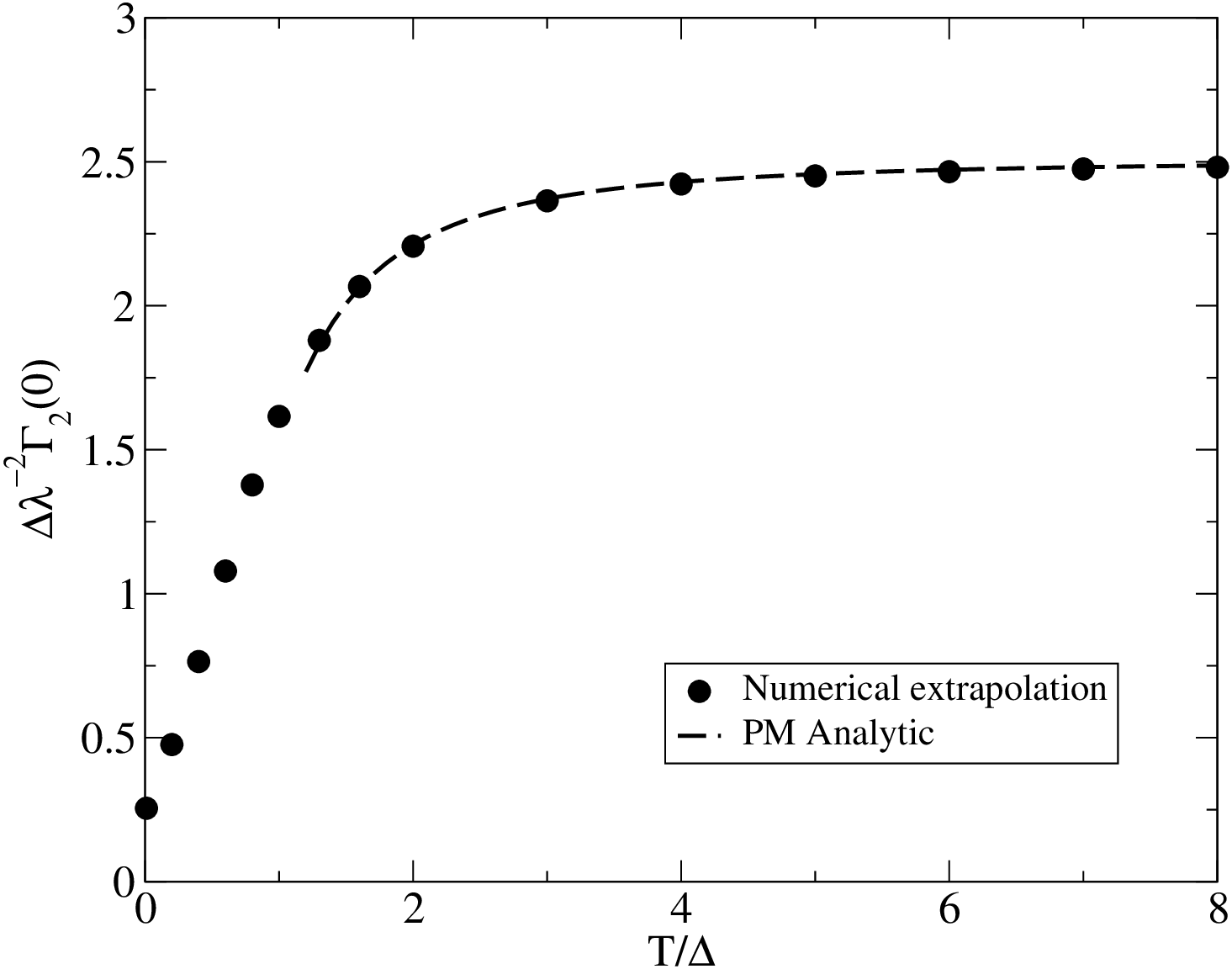}
\caption{\label{fig:csggamma0T}$\Delta\lambda^{-2}\Gamma_2(0)$ (extrapolated to the
thermodynamic limit)
as a function of $T$ for the SK bath.   The dashed curve represents the analytical results in the
paramagnetic phase.}
\end{figure}

The numerical results for  $\ln{M(t)}$  are shown in figure~\ref{fig:csgMt}.  At short times, $t\lesssim t_\Delta\equiv 1/\Delta$, the sum rule obeyed by $\Gamma_2$ ensures that $M(t)$ decays as a gaussian akin to the free spin bath.
For  times $t > {t_\Delta}$, temperature  starts playing a relevant role. At high $T > T_{sg}$, since $\Gamma_2(E)$
is a peak of width $\mathcal{O}(\Delta)$ cf. Fig. \ref{fig:csg97N9}, $M(t)$ decays as a Markovian for all times
$t \sim t_{\Delta}$. 
For  $T \ll T_{sg}$, the asymptotic Markovian regime is preceded by
a power law regime resulting from the partial linear behavior of
$\Gamma_2(E)$ for small $E$.  As $T\to 0$, this intermediate power law regime  is expected to extend to the asymptotic regime. This is however, hard to infer from the numerics, since finite size effects smear the value of \gaz\ and hence, the  linear behavior of \gae\ predicted by the TAP-method calculation [\onlinecite{PP}]. Nevertheless, we see that as $T$ decreases from the paramagnetic phase, \gaz\ decreases and the linear contribution to \gae\ become more significant. 
Extrapolating  the results of $\Gamma_2(0)$ to the limit of $N\rightarrow\infty$ to obtain the Markovian decay rate in the thermodynamic limit, we  find that
$\Gamma_2(0)$ scales  $1/N$ cf. Fig. \ref{fig:csggamma0} (instead of the $1/N^2$ scaling in the Ising chain). The resulting Markovian rate is plotted as a function of $T$ in Fig. \ref{fig:csggamma0T}, and perfectly matches the analytical values predicted in [\onlinecite{CC-prb}]  in the PM-phase. As before,   there is a mismatch of the results in the very low temperature regime, probably because our system sizes do  not access the scaling regime.

As in the case of the Ising chain, we find that
interactions between the spins do lead to longer coherence times as opposed to the case of a free spin bath.
 The only visible effect of the spin glass phase transition is the point of inflexion in the curvature of $\ln M(t)$ for $ t \simeq t_\Delta$ when the asymptotic Markovian regime takes over. This curvature is negative for $T>T_{sg}$ and positive for $T< T_{sg}$. The absence of a radical change in the decoherence is not
surprising since  the qubit does not couple to the operator which corresponds to the
spin glass order parameter.

In the following section, we study the RTIM which is known to have
both, a quantum phase transition at $T=0$, and a finite temperature classical phase transition.
We study two different couplings between the qubit and the bath spin operators and show that the
nature of these couplings has highly non-trivial consequences for the decoherence.

\subsection{Sherrington-Kirkpatrick model in a transverse field $h\neq 0$} \label{sec:TIM}

The Sherrington-Kirkpatrick model in a transverse field (RTIM)  is a more interesting
case than the previously studied $h=0$ case.  At $T=0$, the system undergoes a phase transition from   a spin glass state for  $h\le h_c\simeq1.44\Delta$ to a gapped phase for $h \ge h_c$.
As temperature is increased, the spin glass phase disappears at a finite temperature which depends on the
value of the magnetic field.  The  magnetic field \h\ is therefore, expected to  deeply influence the behavior of \gae, and hence  the coherence of the central spin. Since the model is not isotropic, it is important to note that the spin glass order
exists only along the $x$ component of the spin.  To probe the physical ramifications of this
anisotropy, we examine  two different couplings of the central spin to the bath: 
a coupling of the spin operators in the direction of the field
($a=z$) and a coupling perpendicular to the field ($a=x$). As we will show below, these lead to
radically different predictions for the decoherence.   Previous numerical studies \cite{marcelo-2} of this model  have shown  finite size effects to be  rather minimal. Given the numerical complexities in the
vicinity of the phase transition, we limit the scope of the present study to consider two values of the field $h=0.1h_c$ and $h=2h_c$, that set the system in qualitatively different regimes.

\subsubsection{Coupling parallel  to the field (case $a=z$)} \label{sec:sigmaz}

\begin{figure}[bt]
\centering
\includegraphics[scale=0.25, angle=0]{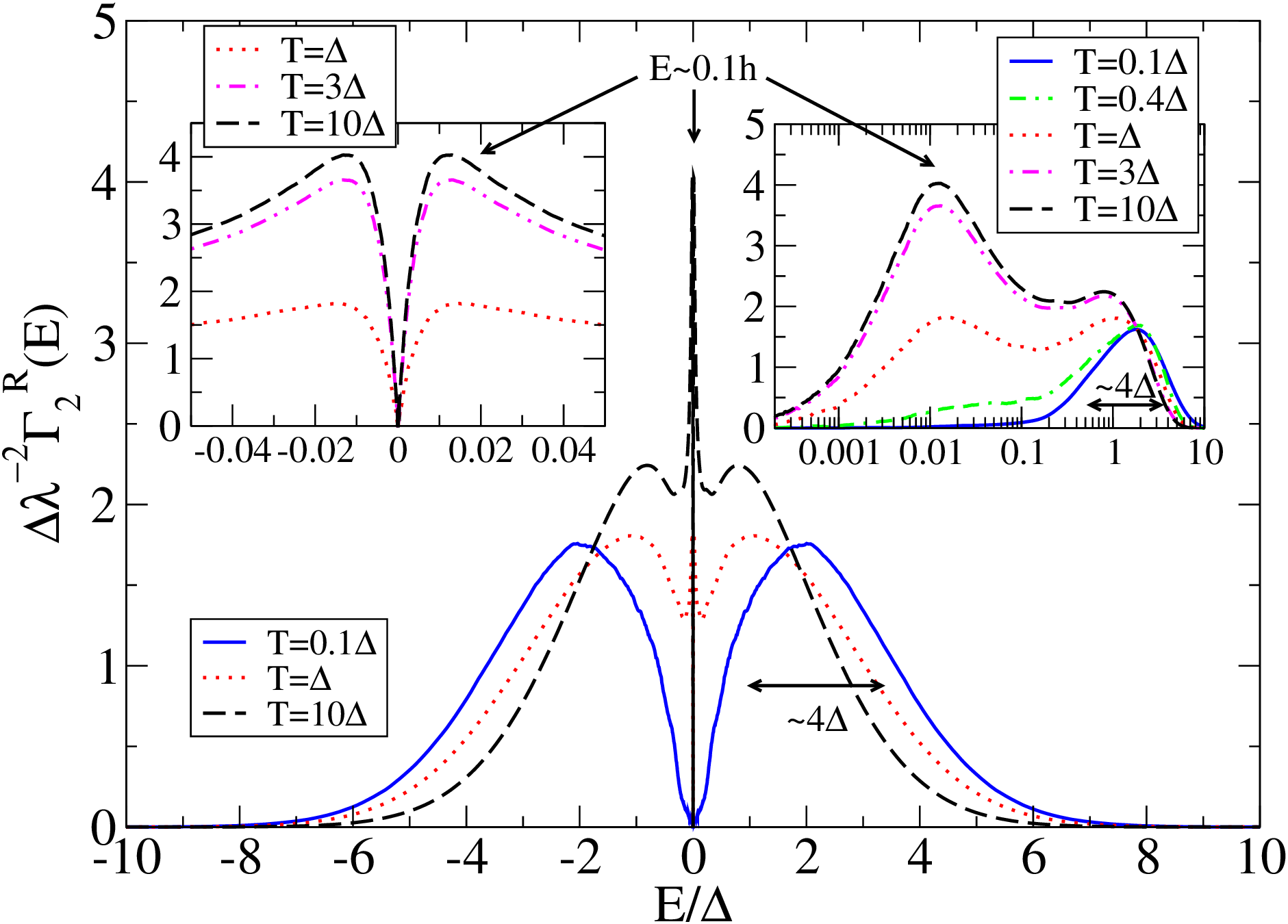}
\caption{\label{fig:csgtrans97N9h0.1hcN9difT}(Color online) $\Delta\lambda^{-2}\Gamma_2(E)$ for different $T$, with  $N=9$ and magnetic field set to \h$=0.1$\hc.
Note the completely 
different behavior from the case $h=0$ (Fig.~\ref{fig:csg97N9}, at small $T$, \gae\ is gapped (right inset, \gae\ in a logarithmical scale).
For higher $T$ (left inset), \gae\ is linear at low energies. We note that the $T=0$ curve is indistinguishable from the one at $T=0.1\Delta$.}
\end{figure}

Since a magnetic moment is present in the $z$ direction for $h \neq0$,  the  \gae\  is given
by the full connected correlation function \eqref{eq:conncorr}.  As shown in (\ref{eq:spectral}), this contributes   a singular term
$\propto \mu\delta(E)$  to \gae\, which then  leads to oscillations in $M(t)$ (equation \ref{eq:mt-osc}). An unambiguous way to extract the coefficient $\mu$  is via the sum rule satisfied by the dynamical
structure factor \cite{marcelo-2}. This, however, is numerically cumbersome and  in the rest of the paper,
we do not deal explicitly with these singular terms, since they only induce oscillations and instead concentrate on 
the non singular part of \gae\ that leads to a decay of the coherence.  
The $T=0$ behavior of \gae\ when a small field \h$=0.1$\hc\ is applied is shown in figure~\ref{fig:csgtrans97N9h0.1hcN9difT}. We note that even for such small fields, \gae\ is radically different from the earlier case of $h=0$. 

\begin{figure}[bt]
\centering
\includegraphics[scale=0.28,angle=0]{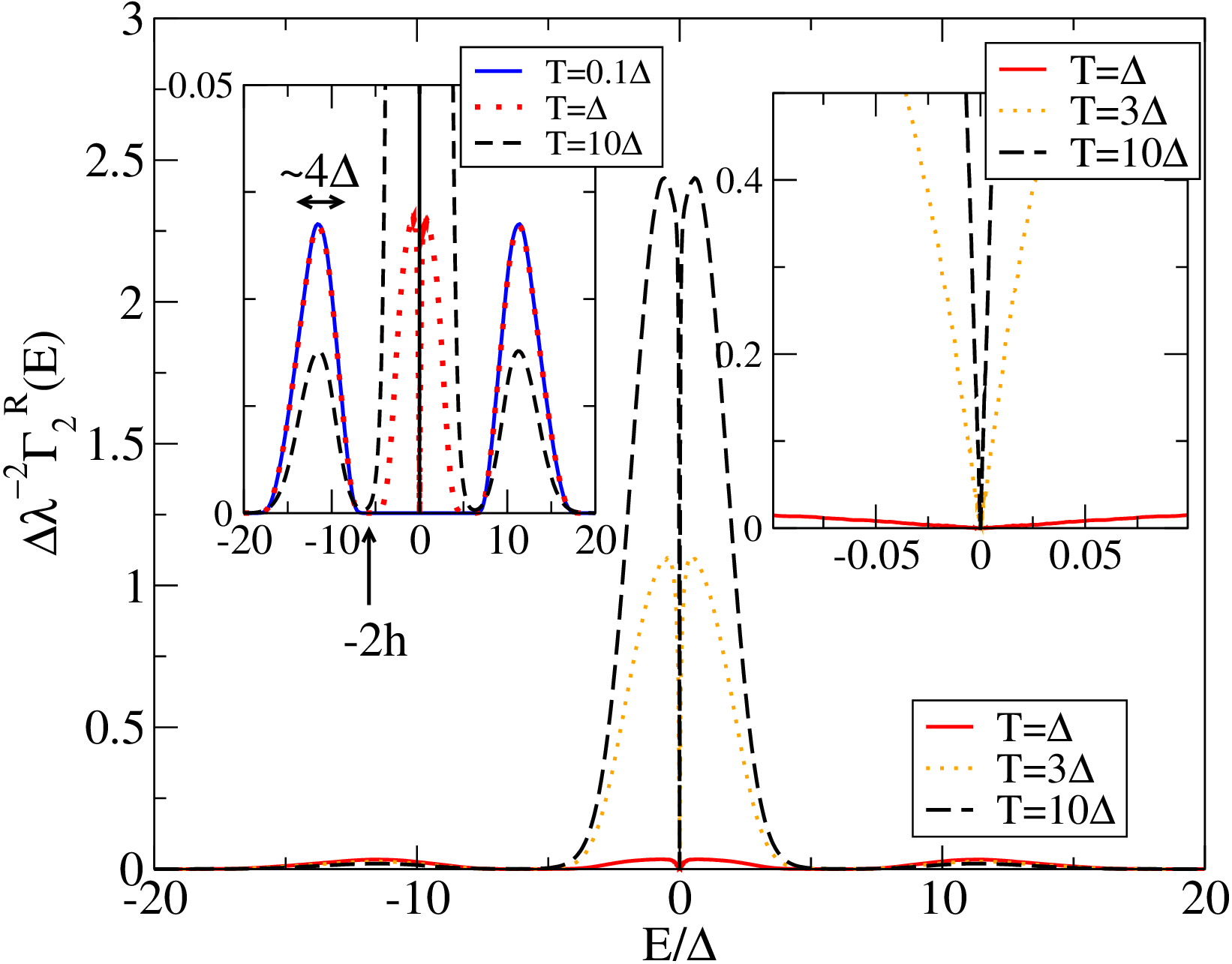}
\caption{\label{fig:csgtrans97h2hcN9difT}(Color online) $\Delta\lambda^{-2}\Gamma_2(E)$ for different 
$T$, with fixed $N=9$ and field \h$=2$\hc (the $T=0$ curve is not plotted here since is indistinguishable from that of $T=0.1\Delta$). Right inset: behavior for  low 
$E$.  Left inset: high frequency peaks. The missing spectral weight is carried by the $\delta(E)$ term
described in the text.}
\end{figure}

Our results for \gae\  obtained for  $N=9$  are shown in Figs.~\ref{fig:csgtrans97N9h0.1hcN9difT}
and~\ref{fig:csgtrans97h2hcN9difT}, for the small and large field case respectively. 
In the former, with $h=0.1h_c$, \gae\ exhibits a gap of order
$2h$ at $T=0$ (numerically, the $T=0$ curve is indistinguishable from the $T=0.1\Delta$ curve shown in 
Fig.~\ref{fig:csgtrans97N9h0.1hcN9difT}).  
At high $T$, \gae\ shows a pronounced peak around $E\simeq\pm0.1h$ and broader peaks 
of width $\approx 4 \Delta$ in the background.
Similar features are seen at high magnetic fields (Fig.~\ref{fig:csgtrans97h2hcN9difT}): existence of a gap   
$2h$ at $T=0$ and a two peak structure of width $4\Delta$ at high temperatures. For high enough $T$,
the susceptibility at low frequencies becomes linear with a 
temperature dependent slope, $\Gamma(E) =\alpha(T)  |E|$ as $E \to 0$.
Note however, that in the high field case the $\delta(E)$ term arising from moment formation
carries most of the spectral weight at $T=0$, as the spectral weight of  
the regular part of \gae\  decreases dramatically. The principal difference between the
high and low field case is the presence of the sharp peak around $E\sim 2h$ in the
low field case. We have studied the finite size scaling of \gae\ and we find that the features mentioned
above are robust to 
finite size effects. Moreover,  contrary to the previous cases, here we obtain a good
convergence with system size in the low frequency regime. 

Interestingly,  our results bely the naive expectation  that the behavior in the high temperature   paramagnetic phase  be 
 independent of the values of \h\ considered here. 
This can be attributed to the fundamental difference between the structure of eigenfunctions and eigenvalues in the two
cases: for $h=0.1h_c$, they correspond to that of
a spinglass system, while the high field case essentially correspond to spins in a strong field
where the interactions between the spins can be viewed as a perturbation. 
At high enough  temperatures, all eigenfunctions and eigenvalues contribute equally  to the \gae\, thus 
the large $T$ regime actually reveals the ``geometric'' underlying structure of the Hamiltonian, rather
than a naively expected universal paramagnetic regime of free spins.
This observation remains relevant for all the different spin-models analyzed in this work.

\begin{figure}[bt]
\centering
\includegraphics[scale=0.32,angle=0]{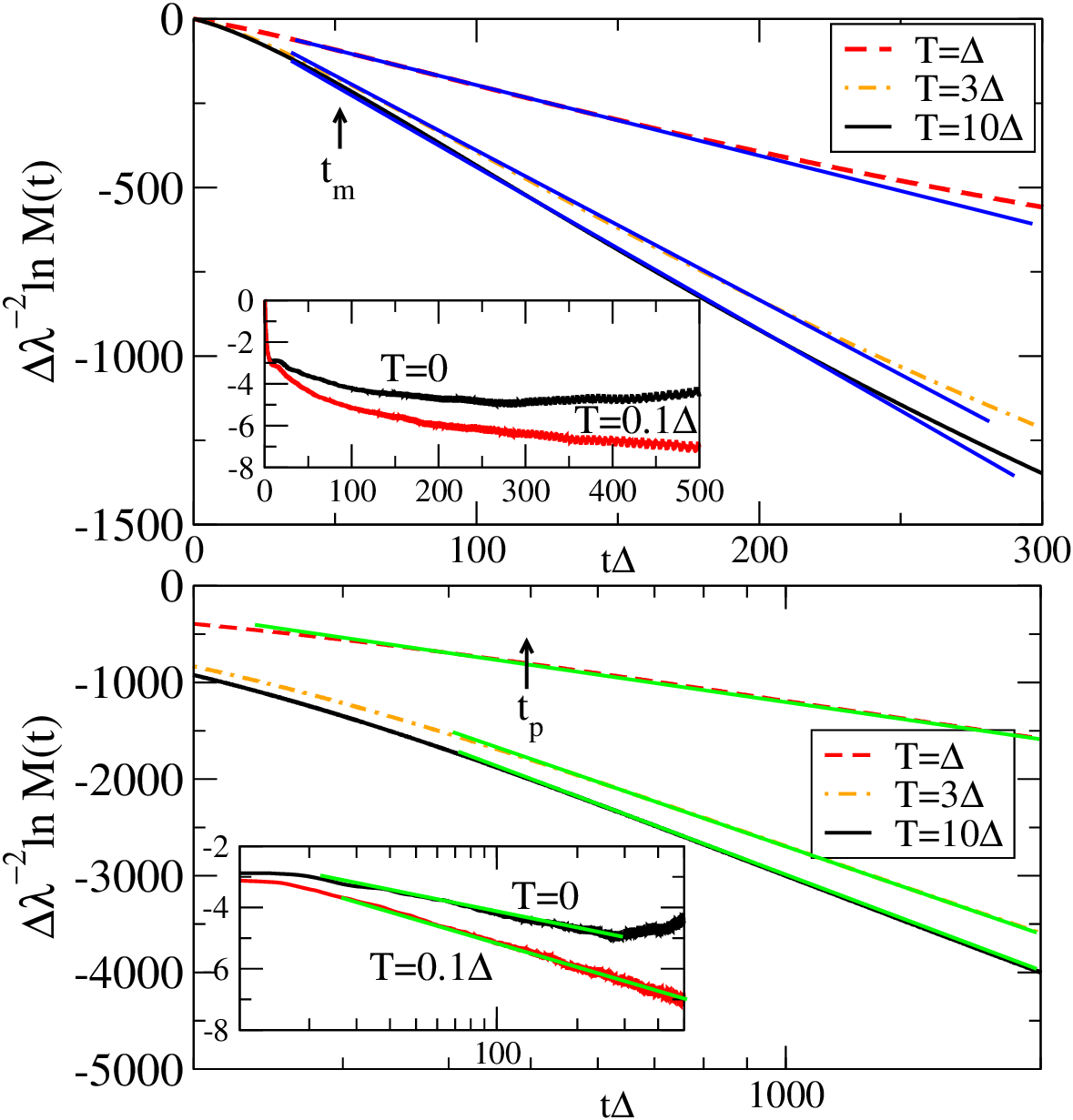}
\caption{\label{fig:csgtrans70h0.1hctodasT}(Color online) $(\Delta/\lambda)^{2}\ln{M(t)}$ with
$h=0.1$\hc. The upper graph (linear scale), shows an intermediate time 
Markovian regime ( $t > t\simeq{t_m}$), while the lower graph (logarithmic scale)
 shows a final power law regime $t > t\simeq{t_{p}}$). Insets:
$\ln{M(t)}$ for low $T$.}
\end{figure}

\begin{figure}[bt]
\centering
\includegraphics[scale=0.32,angle=0]{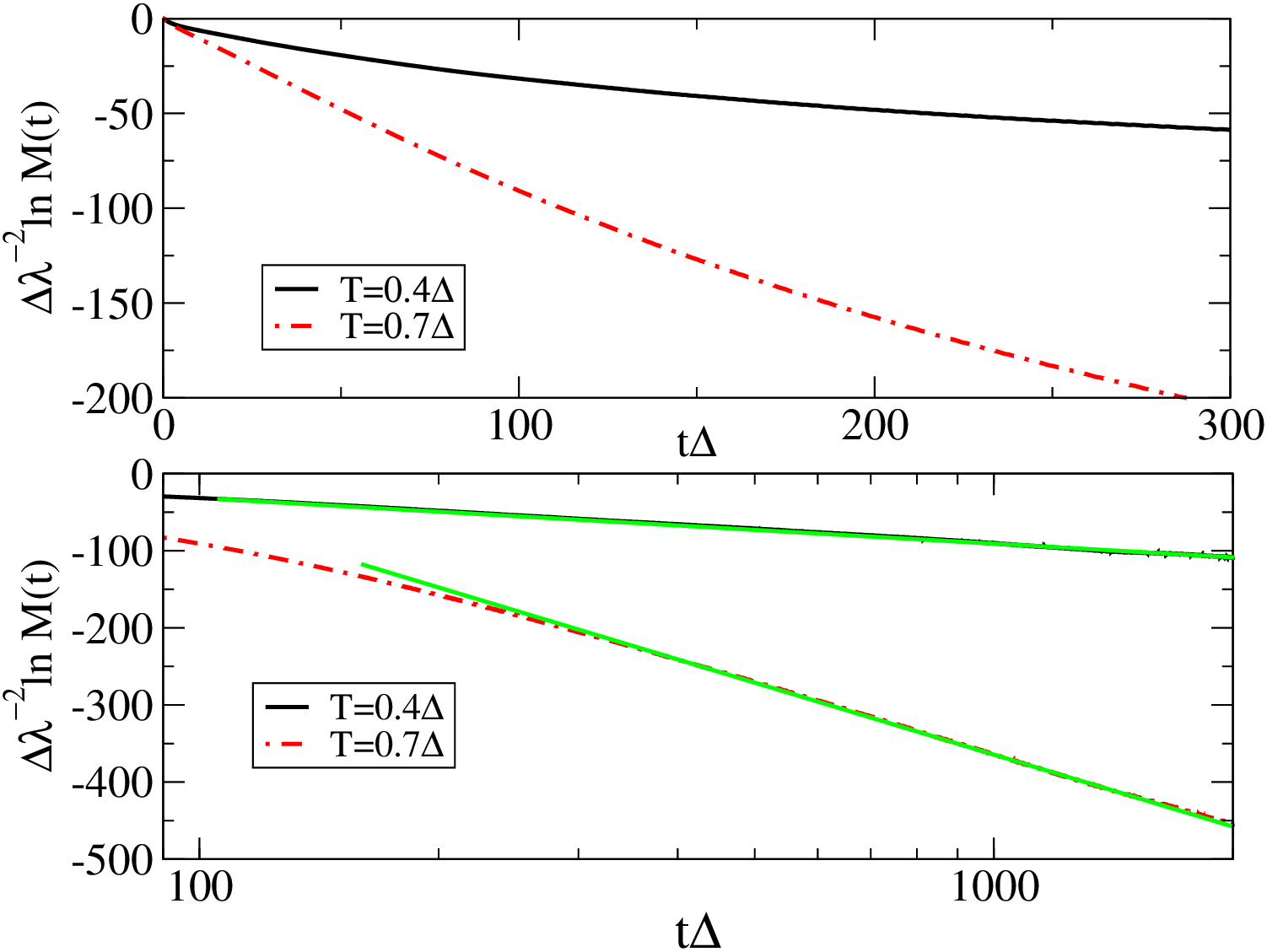}
\caption{\label{fig:csgtrans70h0.1hcT04y07}(Color online) $(\Delta/\lambda)^{2}\ln{M(t)}$ with
$h=0.1$\hc. The upper graph (linear scale), in contrary to the high $T$ coherence of Fig. \ref{fig:csgtrans70h0.1hctodasT}, does not show an intermediate Markovian regime, while the lower graph (logarithmic scale) shows a final power law regime in which $t_{p}$ increases with $T$.}
\end{figure}

The results for \gae\ indicate a very rich evolution of the coherence of the central spin, particularly 
at long times. We first analyze the results for $M(t)$ plotted in
Figs.~\ref{fig:csgtrans70h0.1hctodasT} (high $T$) and \ref{fig:csgtrans70h0.1hcT04y07} (low $T$), for $h=0.1h_c$.
As before, for $t\lesssim$\td,  the sum rule obeyed by \gae\ \eqref{eq:sumrule}  dictates an universal gaussian 
decay for the decoherence, independently of \h, $\Delta$ and $T$. This is indeed
verified by our results.


\begin{figure}[bt]
\centering
\includegraphics[scale=0.25,angle=0]{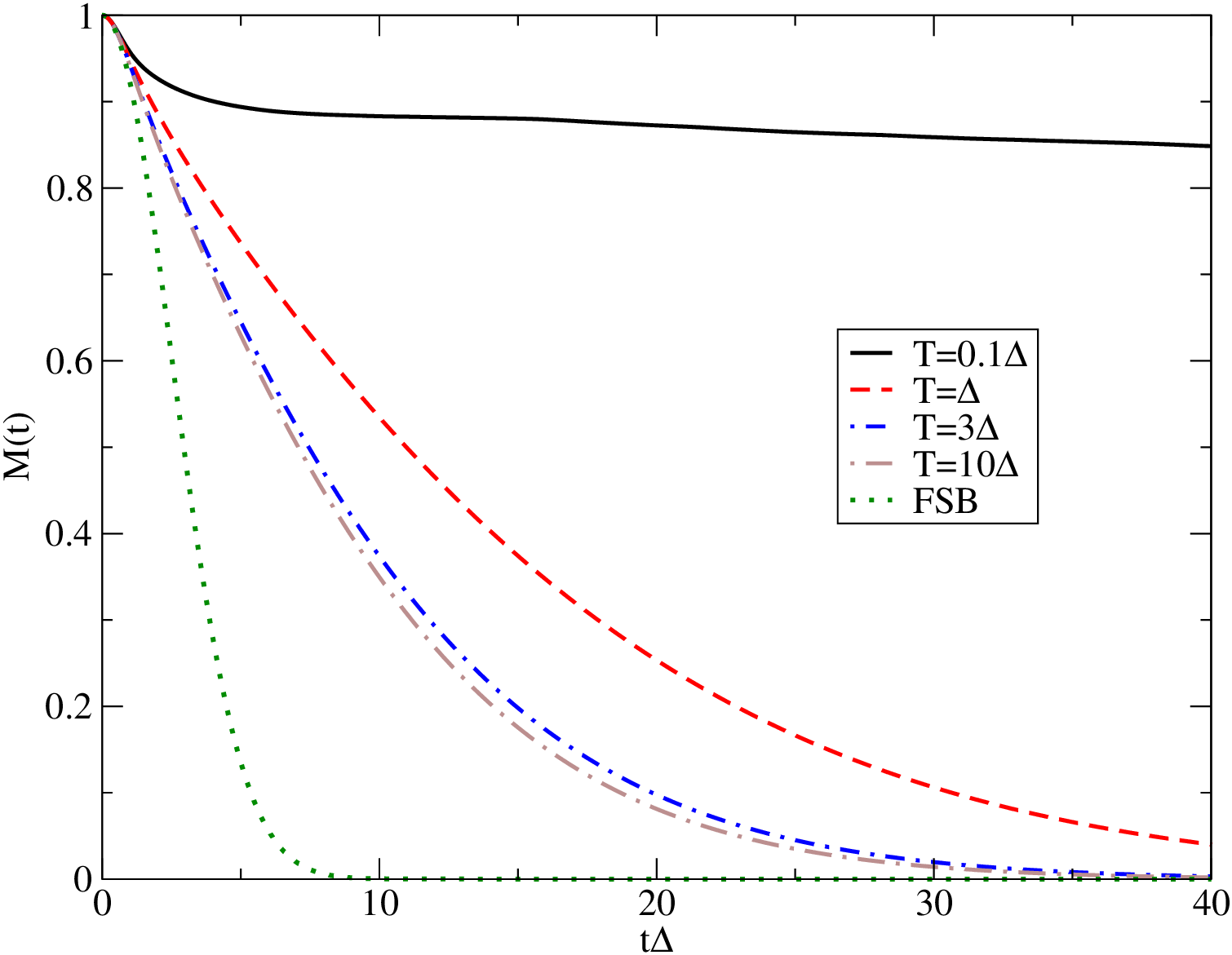}
\caption{\label{fig:csgtransMth0.1hcdifTN10z}(Color online) Coherence of the central spin as a 
function of  time, for a bath of $N=10$ spins for $h=0.1h_c$ and a bath-qubit coupling 
 $\lambda=0.2\Delta$. The free spin case (FSB) is also plotted for reference.}
\end{figure}

For $T=0$ (insets of Fig. \ref{fig:csgtrans70h0.1hctodasT}), the
decoherence is only partial due to the presence of a gap in \gae\
and shows oscillations with a frequency proportional to the size of
the gap. Similar oscillations are also seen at very low temperatures. 
The short time regime $t< t_\Delta$ is followed by a  power-law regime
for $t > t_p$
which saturates to a finite value at long times.
 As $T$ increases, the short time and asymptotic power law regime are
separated by an intermediate Markovian regime 
for $ t_m < t \ll t_{p}$.  For comparison,  the coherence for a bath with $N=10$ spins is plotted in figure \ref{fig:csgtransMth0.1hcdifTN10z} various $T$.  As in the case for $h=0$, not dramatic change is
seen as one traverses the spin glass transition temperature. The existence of an asymptotic power law
regime at finite temperatures completely defies the conventional lore
that the asymptotic decoherence at finite temperatures is Markovian
\cite{pb,lpm}. A systematic analysis of our numerical results
indicates that for low fields, $t_p$ is an increasing function of
temperature. This implies that in the high temperature paramagnetic
phase, the power law regime is pushed to ultra long times and the
decoherence is Markovian for realistic times. Though the exact values of $t_p$ and $t_m$ are subject
to finite size effects, we have studied various system sizes and find
that
the  intermediate Markovian and power law regimes will survive in the
thermodynamic limit.

Before moving on to the high magnetic field case, we shall 
present an intuitive physical picture of the very low field case.
We first recall the physical picture of the spin-glass ground state at $T=0$ and $h=0$, that
was discussed in the numerical investigation of Ref. \onlinecite{marcelo-2}.  
There it was argued that the spin system can be viewed as a coexisting collection of
large non-frustrated clusters of spins with few remnant strongly frustrated ``dangling'' spins.
Thus, the dangling spins experience a distribution of effective magnetic 
fields $h_{\textrm{eff}}$ which, from the work of
Sherrington and Kirkpatrick is known to have a linear distribution.
This linear is directly reflected in the \gae\ $\sim |E|$ that we discussed before.

Now, when a small external magnetic field is turned on, the unfrustrated 
spins of the clusters will remain essentially unaffected.
In contrast, the dangling spins will align in the direction of $\textrm{max}(h_{\textrm{eff}},h)$. 
This means that the finite external field $h$ will act as a low frequency cut-off, pushing spectral weight towards higher
frequencies and thus opening an $h$-controlled gap in \gae\ around $E=0$
(see the low-$T$ curve of the right inset of Fig. \ref{fig:csgtrans97N9h0.1hcN9difT}). 
This behavior is in stark contrast with the reference case $h=0$ presented in section \ref{sec:results} where \gae\ remains
always gapless (Fig. \ref{fig:csg97N9}). The dramatic change in the low $E$ regime implies also the qualitative
change in the long time behavior of the decoherence that we described in the previous section. 

\begin{figure}[bt]
\centering
\includegraphics[scale=0.3,angle=0]{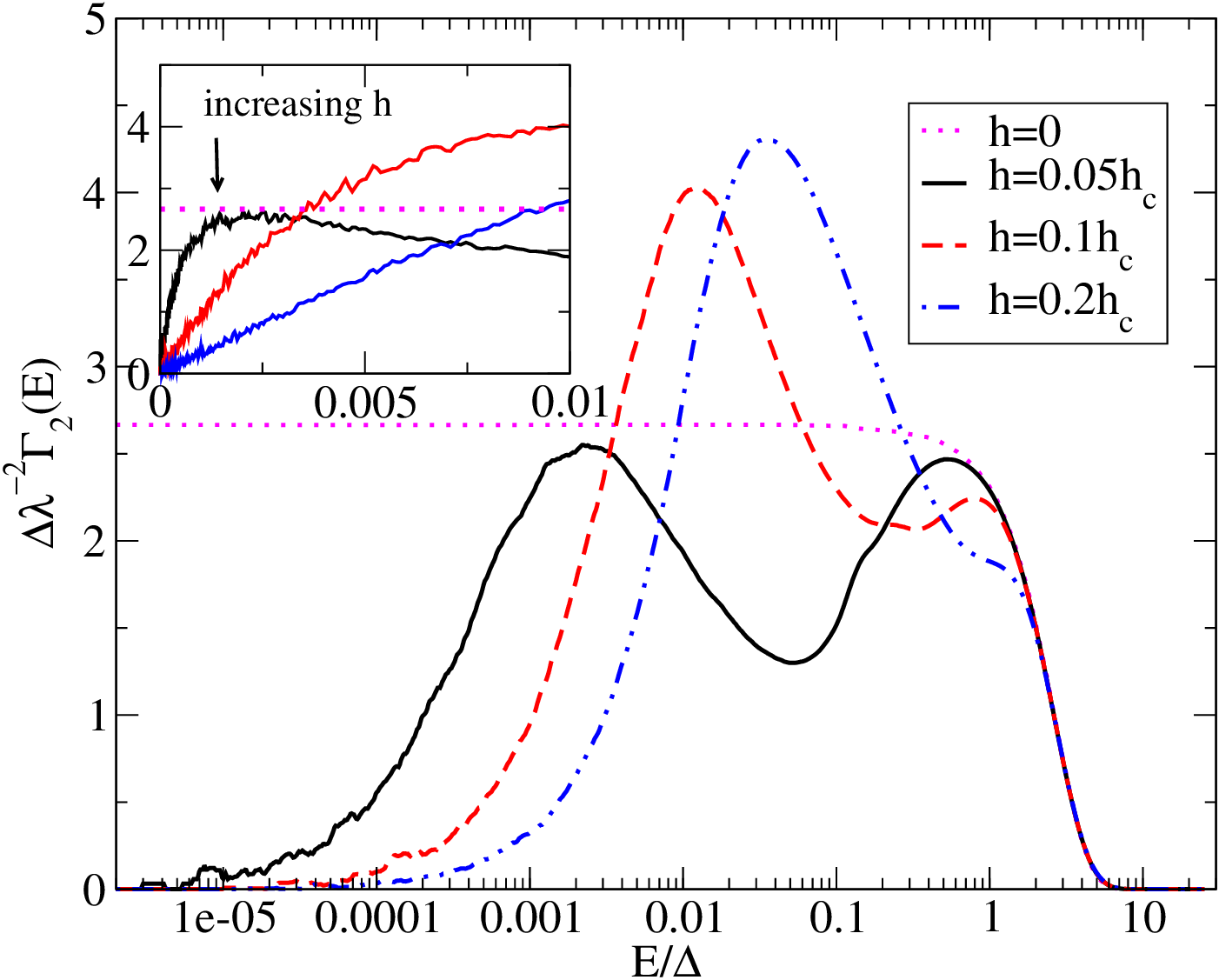}
\caption{\label{fig:csgtransdifh}(Color online) $\Delta\lambda^{-2}\Gamma_2(E)$  as a function of $E$
for different values of the field \h\, with $N=9$ and $T=10\Delta$. Inset:  linear  behavior of \gae\  for low energies.}
\end{figure}

This line of argumentation also allows us to qualitatively understand the origin
of the linearity of \gae\ at high $T$.
In this limit,  from (\ref{eq:spectral}) we see that all eigenvalues and eigenfunctions contribute to 
\gae. On the other hand,
when $h=0$, the data of Fig. \ref{fig:csg97N9} shows that \gae\ is finite as $E \to 0$, so 
there are plenty of contributions at arbitrary low frequencies, which originate in pairs of quasi-degenerate
eigenstates $|n\rangle$ and $|m\rangle$ with eigenenergies $E_n \approx E_m$, which also have a non-vanishing matrix 
element $s^z_{nm} = \langle n|\sigma^z_i|m\rangle$.
When a small external field ( \h\ $\ll  $ \hc\ ) is turned on we may consider it as a perturbation an perform the
following qualitative analysis: 
The (2x2) block of $H$ in the subspace of $n$ and $m$ will read:
\begin{equation*}
 H_{nm} = \left(
  \begin{array}{ccc}
  E_o & h s^z_{nm} \\
  h s^z_{nm} & E_o
  \end{array} \right)
\end{equation*}
where $E_o$ is the quasi-degenerate energy of the states.
The effect of the small $h$ is to lift the degeneracy of the pair of levels 
resulting in a transfer of spectral weight from $E \approx 0$ to higher frequencies.
Extending the analysis to all pairs of quasi-degenerate states contributing to
\gae\  implies the immediate collapse of the finite value of $\Gamma_2(E=0)$ to zero when \h$\not=0$,
as is seen in our results of Fig. \ref{fig:csgtransdifh}.
Moreover, the new eigenvalues of $H_{nm}$ are $\approx \pm h s^z_{nm}$, thus with the reasonable assumption of
a featureless distribution of the $s^z_{nm}$ matrix elements, we conclude that the degeneracy lifting is uniformly
distributed and directly proportional to $h$. This implies that the behavior of \gae\ at very low frequencies
should be linear in $E$ with a slope roughly given by $1/h$. This analysis is in qualitative agreement with the low field
data shown in Fig. \ref{fig:csgtransdifh}.

\begin{figure}[bt]
\centering
\includegraphics[scale=0.25,angle=0]{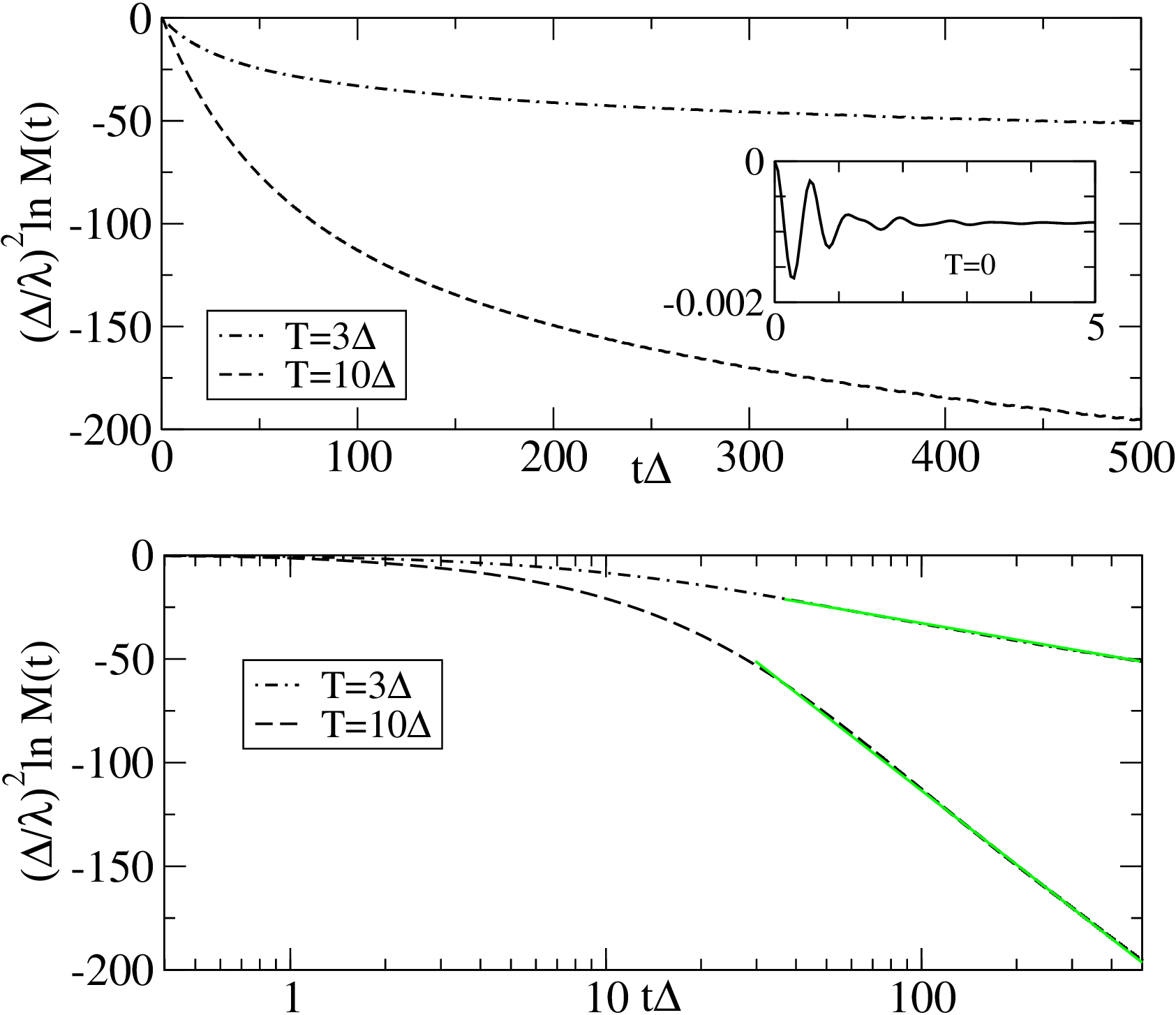}
\caption{\label{fig:csgtrans70h2hctodasT}(Color online) $(\Delta/\lambda)^{2}\ln{M(t)}$ as a function of time for \h$=2$\hc. For low $T$, since \gae\ is gaped, the decoherence is very weak and shows oscillations (inset). As $T$ increases, the coherence decays asymptotically as a power law.}
\end{figure}

\begin{figure}[bt]
\centering
\includegraphics[scale=0.32,angle=0]{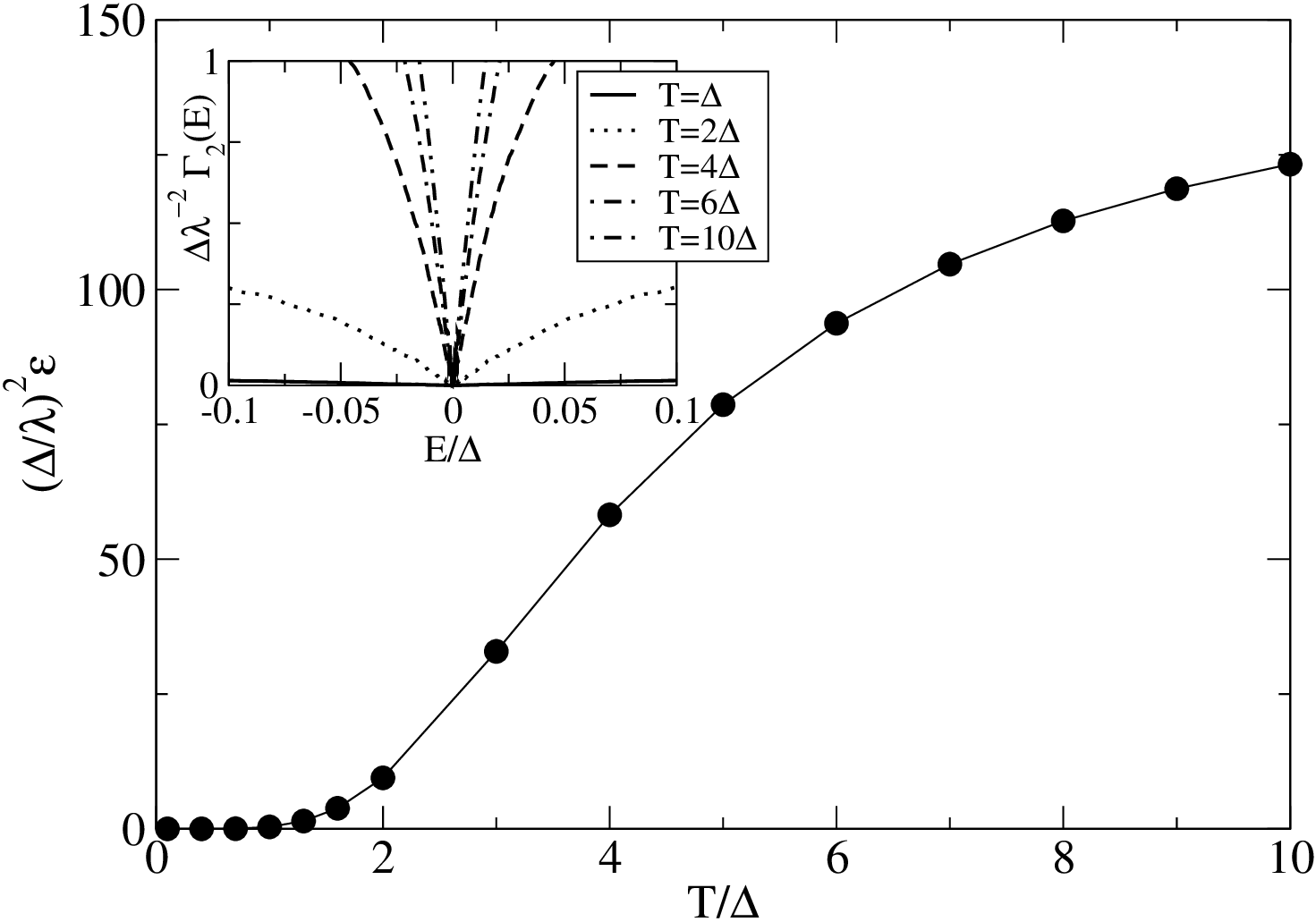}
\caption{\label{fig:csgtransslopeTh2hc}The exponent $\epsilon$ ($M(t)\propto t^{-\epsilon}$) extrapolated to the thermodynamic limit, as a function 
of $T$ for $h=2 h_c$. }
\end{figure}

At high magnetic fields (Fig.~\ref{fig:csgtrans70h2hctodasT}), the coherence 
presents a different qualitative behavior. At low $T$, since \gae\ is
nearly   gapped, and most of the spectral 
weight is in the $\delta(E)$ term, the central spin decoherence is
very weak and shows  oscillations. At high $T$, in contrast to the low \h\
behavior, there is no well defined intermediate Markovian regime,
but we recover a  power-law decay ($M(t)\propto t^{-\epsilon}$) at
asymptotic times. This coefficient $\epsilon$, which characterizes the power law decay, increases 
with $T$ and saturates to a finite value at very high temperatures, proving that the  qubit decoheres faster as  temperature increases.  The value of the exponent $\epsilon$ extrapolated  to 
the thermodynamic limit, as a function of $T$, is shown in
figure~\ref{fig:csgtransslopeTh2hc}.

\subsubsection{Coupling transverse to the field (case $a=x$)}

We now explore the case  where  the qubit  couples directly to the bath operator  which
is related to the spin glass order parameter. In the spin-glass part of the phase diagram ($h \leq h_c$),
we now have
  $\overline{\langle \sigma_x^i \rangle} = 0$ and   $\overline{\langle \sigma_x^i \rangle}^2 \neq 0$ 
$\forall i$.  As in  Sec. \ref{sec:sigmaz}, \gae\ is again a  sum of a regular and  a  singular contribution. 
The singular part is a  $\delta(E)$ contribution whose strength is related to the spin glass order parameter or Edwards Anderson parameter \qea,
$\Gamma_2^{\textrm{NR}}=(4\pi$\qea$ - \sum_i {\overline{\langle\sigma_x^i\rangle^2}} )\delta(E)$.
The non-regular part induces oscillations in $M(t)$ and the regular part results in the decay  of $M(t)$ \eqref{eq:mtilde}. 
The  estimation of \qea\ is a central problem in spin-glasses and as expected our numerics do not produce well converged results
of the values of this parameter. 
 \qea\ is typically estimated by systematically studying  \gae\ for 
different sizes of the system, as in Ref. \onlinecite{marcelo-2}. Though it is reasonably straightforward to estimate  \qea\ at $T=0$,  this is not the case  at finite temperatures where  the delta peak is broadened by both  finite size  effects and thermal excitations.\cite{marcelo-2} 

\begin{figure}[bt]
\centering
\includegraphics[scale=0.35,angle=0]{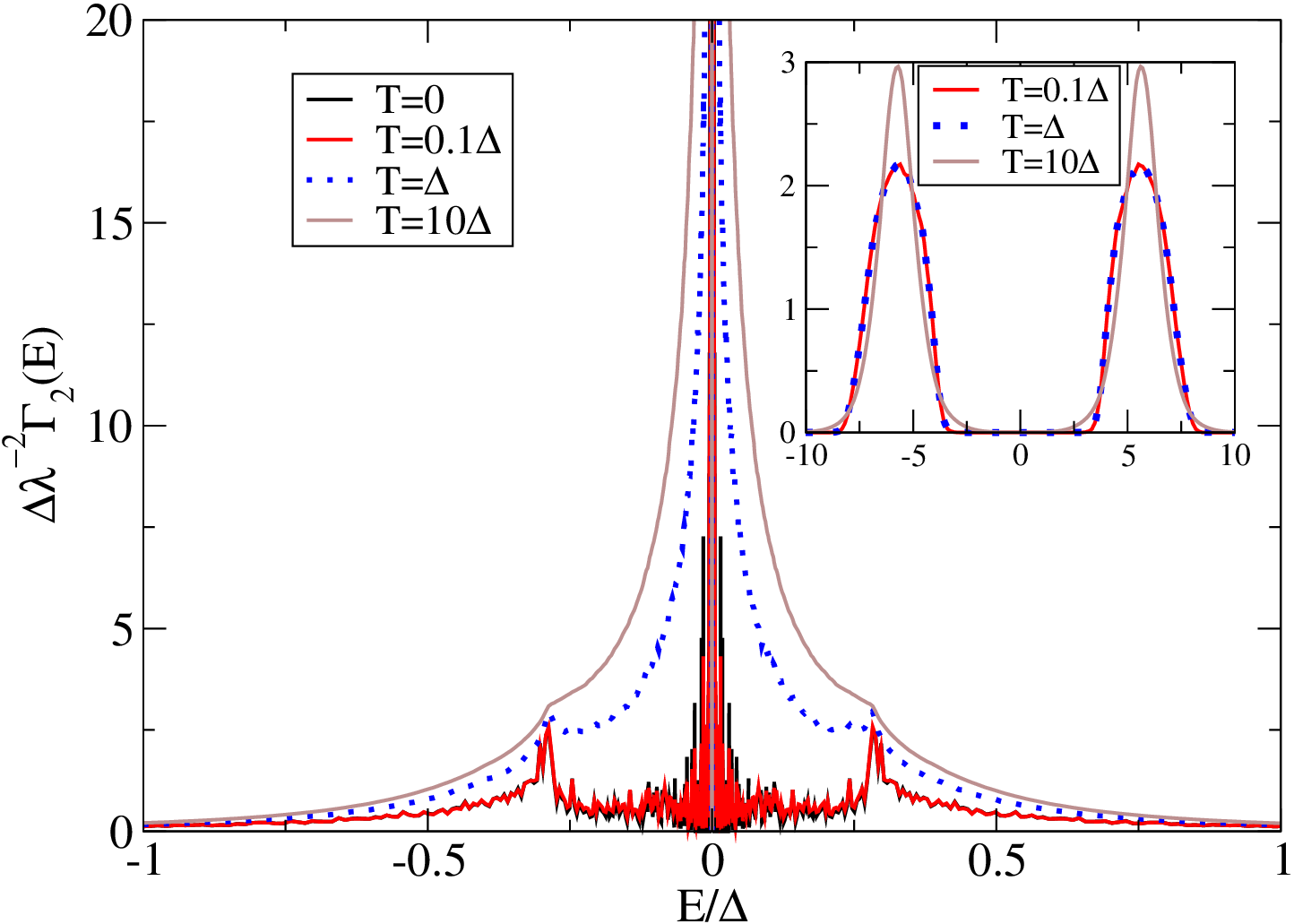}
\caption{\label{fig:csgtrans87h0.1hcN9difT}(Color online) $\Delta\lambda^{-2}\Gamma_2(E)$ 
for N=9  and field \h$=0.1$\hc for different $T$. For $T=0$, \gae\ increase 
linearly for low $E$ and has a maximum at $E=2h$. As $T$ increases, 
spectral weight from the $\delta(E)$ (discussed in text)  is transferred to low energies. Inset: \gae\ for
$h=2$\hc\ showing thermal smearing of the gap.}
\end{figure}

When $h=0$, the operator $V^x$ is constant of motion since  $[H_B,V^x]=0$ 
resulting in a temperature independent  $\Gamma(E) = 4\pi\lambda^2\Delta^{-1}\delta(E)$. As discussed
in Sec. \ref{sec:formalism}, this results in  a purely oscillatory behavior of the form
$M(t) = \cos( 2\lambda\Delta^{-1} t)$.
When  the field \h\ is turned on, spin flips are allowed, $[H_B,V^x] \neq 0$ and  \gae\ has a richer structure.
For small $h$ ($h\ll h_c$) and $T=0$, in addition to a $\delta(E)$ term \gae\ has a regular part
 $\Gamma^{\textrm{R}}(E) \propto |E|$  as $E \to 0$ and has a  peak  of width $4\Delta$, centered around  $E=2h$\cite{marcelo-2}.
As temperature is increased, since \qea\  is expected to decrease,  a part of the spectral weight of the  $\delta(E)$ term is transferred 
 to the regular part $\Gamma^{\textrm{R}}$ which is now nonlinear  at low energies (Fig.~\ref{fig:csgtrans87h0.1hcN9difT}). 
 
We can now construct a picture of the coherence in the spin-glass phase based on the
above description of \gae. The short time behavior is governed by  a gaussian.  Due to the afore-mentioned problem with the broadening of the
delta term, we have not been able to obtain clear predictions for the intermediate and asymptotic
decoherence  for low \h\ and low $T$.
This is intimately linked to the underlying spin-glass order since
in this case some of the contribution to the singular part exists only in the
thermodynamic limit, as opposed to the preceding section, where the singular
term arises from a straightforward magnetization of the underlying system and hence
one did not have to deal with broadening induced by finite size. 
 Nonetheless, for 
$T=0$ and $h\ll h_c$  it is possible to have a qualitative picture of the coherence.  As in the zero-field case,  the linearity of \gae\ as $E \to 0$ leads to a power-law asymptotic 
decay of the coherence. At finite temperatures,  we expect the asymptotic decay to be Markovian.

\begin{figure}[tb]
\centering
\includegraphics[scale=0.3,angle=0]{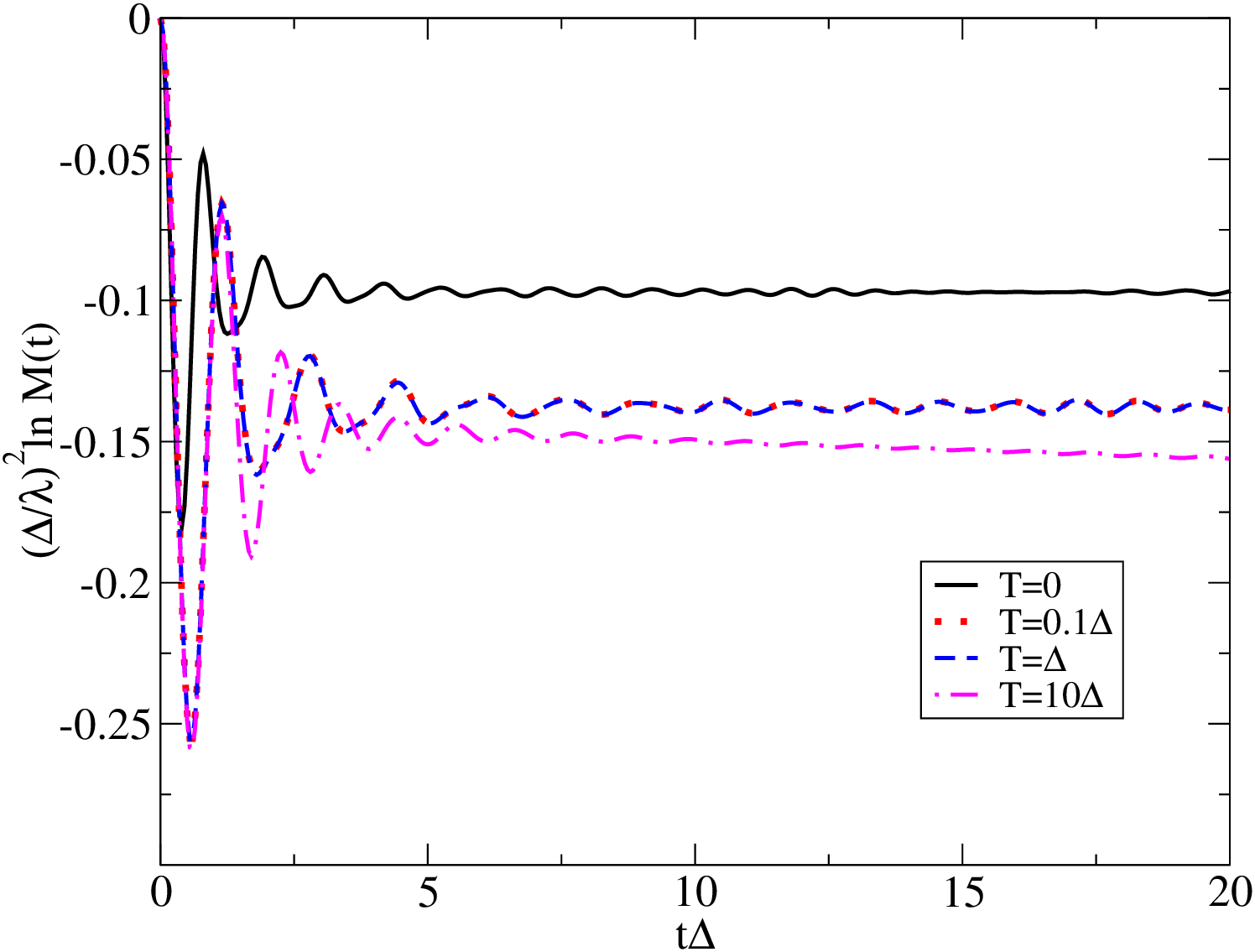}
\caption{\label{fig:csgtrans80h2hcN9difT}(Color online) $\Delta\lambda^{-2}\ln M(t)$ 
for N=9 and field \h$=2$\hc\ for different $T$. For $T=0$, since \gae\ is gapped, the decoherence
is partial but asymptotic Markovian decoherence is recovered at finite  $T$.}
\end{figure}

For high \h\ ($h>h_c$), there is no
spin glass order and hence no singular contribution to \gae.
At $T=0$, the function \gae\ is gaped, and it is 
centered in $E=\pm 2h_c$, with a width of $\sim 4\Delta$. As $T$ is increased, thermal 
excitations appear within the gap 
(Fig.~\ref{fig:csgtrans87h0.1hcN9difT}). Thus, in the former case the decoherence is partial
 (Fig. \ref{fig:csgtrans80h2hcN9difT}) and we see
oscillations with a frequency given by the gap size.  As $T$ increases, thermal excitations
start to fill the gap and  $M(t)$  decays in a 
Markovian way given by the value of \gaz. This behavior conforms to the usual  expectations of asymptotic Markovian decay at
finite temperatures and is very different from the small $h$ case.

\section{Conclusions} \label{sec:conclusions}

In this paper, we have used exact diagonalization methods to  study the decoherence  of a central spin induced by
spinbaths with random interactions. We find that   the asymptotic decoherence is intricately linked with the
nature of the interactions in the bath. Moreover, for a given set of bath parameters, the decoherence
crucially depends on the nature of the coupling  between the qubit and the bath spins.
For the cases of the random transverse
Ising model studied here, we find that the underlying nature of the eigenstates
of the bath hamiltonian play a preponderant role in determining the decoherence
even in the finite temperature paramagnetic phase.  More precisely, we find that  the decoherence
in some of the finite temperature paramagnetic phases  is strongly non-Markovian.
We emphasize that
standard Markovian approximations  used  to obtain the density matrix and the decoherence at
finite temperatures  should not be used blindly as they can lead to highly misleading results as
in the present case of disordered interacting systems.  Unfortunately, we have not been able
to study the decoherence in the vicinity of  the phase transitions in the bath ( both at zero and finite temperatures).  This requires the use of other numerical and analytical methods which is beyond the
scope of the present work.  It will be interesting to study if these
features are seen in other highly frustrated spin baths and whether this non-Markovian behavior
survives when order corrections to the self-energy are taken into account.   These questions are
left for future work.

We would like to thank S. Camalet for interesting discussions. One of us R.C. acknowledges support
from the Institut Universitaire de France. 

\end{document}